\documentclass[prd,twocolumn,reprint,preprintnumbers,nofootinbib,superscriptaddress,longbibliography]{revtex4-1}

\usepackage{slashed,color}
\usepackage{dcolumn}
\usepackage{enumitem}

\usepackage{bm}
\usepackage{amsmath} 
\allowdisplaybreaks
\usepackage{amssymb}  
\usepackage{enumerate}
\usepackage{multirow}
\usepackage{braket}
\usepackage[toc,page]{appendix}
\usepackage{url}
\usepackage{graphicx}
\usepackage{subcaption}
\usepackage{hyperref}
\hypersetup{colorlinks=true}
\usepackage{latexsym}
\usepackage{wasysym}
\usepackage{dhucs-enumerate}
\usepackage[most]{tcolorbox}

\usepackage{setspace}
\usepackage[skip=2pt]{caption} 
\usepackage{mathrsfs,amssymb}  
\usepackage{cancel}
\usepackage[normalem]{ulem}
\usepackage{cleveref}
\usepackage{tikz}
\usetikzlibrary{decorations.markings}
\usetikzlibrary{positioning}
\usetikzlibrary{shapes}
\usetikzlibrary{calc}
\usepackage{textcomp}
\usepackage{lipsum}
\usepackage{newtxtext, newtxmath}
\usepackage[export]{adjustbox}
\usepackage{tabularray}
\UseTblrLibrary{booktabs}
\usepackage{float} 

\usepackage{caption}
\usepackage[braket, qm]{qcircuit}

\usepackage{hhline}
\usepackage{soul}

\begin{document}


\title{
High-Frequency Gravitational Wave Detection with Superconducting Qubits
}

\author{Heechan Yi}
\email{lhc320@yonsei.ac.kr}
\affiliation{Department of Physics, Yonsei University, Seoul 03722, Republic of Korea}

\author{Seong Chan Park}
\email{sc.park@yonsei.ac.kr}
\affiliation{Department of Physics, Yonsei University, Seoul 03722, Republic of Korea}
\affiliation{Korea Institute for Advanced Study, Seoul, 02455, Korea}
 
\author{Kyoungchul Kong}
\email{kckong@ku.edu}
\affiliation{Department of Physics and Astronomy, University of Kansas, Lawrence, KS 66045, USA}
 
\author{Myeonghun Park}
\email{parc.seoultech@seoultech.ac.kr}
\affiliation{School of Natural Sciences, Seoultech, Seoul 01811, Republic of Korea}
\affiliation{Institute of Convergent Fundamental Studies, Seoultech, Seoul 01811, South Korea}
\affiliation{Institute for Basic Science, Daejeon 34126, South Korea}%


\date{\today}

\begin{abstract}
High-frequency gravitational waves (HFGWs) provide a unique window into high-energy and early-universe physics, yet they evade traditional macroscopic interferometry. To bridge this detection gap, we propose a novel quantum-sensing paradigm utilizing superconducting transmon qubits embedded in resonant microwave cavities. Through the inverse Gertsenshtein effect, HFGWs propagating in a static magnetic field resonantly excite a cavity mode. By leveraging the characteristic spin-2 quadrupolar pattern of the induced electromagnetic field, we position qubits directly at the electric-field hot spots of the $\mathrm{TE}_{212}$ mode to act as localized sensors. Crucially, configuring this array as an entangled quantum register via symmetric Dicke states unlocks a fundamental scaling advantage: the signal probability scales quadratically with the qubit number, translating to a $h_{\min} \propto n_q^{-3/4}$ strain sensitivity scaling. We demonstrate that an idealized global register of 800 qubits reaches a strain sensitivity that surpasses standard macroscopic cavity-power limits by five orders of magnitude. Benchmarked against representative axion-haloscope parameters, this collective quantum enhancement decisively mitigates the profound Planck-scale suppression inherent to gravitational interactions, establishing a transformative framework for next-generation HFGW searches in the GHz band.
\end{abstract}


\maketitle

\section{Introduction}
\label{sec:introduction}

The historic first detections of gravitational waves (GWs)~\cite{LIGOScientific:2016aoc} opened an entirely new observational window into the Universe. The signals observed to date lie within the $\mathrm{Hz}$ to $\mathrm{kHz}$ band, defining the operational envelope of current ground-based macroscopic interferometers. Much like the electromagnetic spectrum, however, GWs are expected to span a vastly broader range of frequencies, with each band encoding distinct information about its astrophysical or cosmological origins~\cite{longair2011high}. To probe frequencies beyond the reach of current facilities, several advanced detector architectures have been proposed, including future space-based interferometers~\cite{LISA:2017pwj}, atom interferometers~\cite{MAGIS-100:2021etm, Bertoldi:2021rqk, Badurina:2019hst}, pulsar timing arrays~\cite{NANOGrav:2020bcs}, and cosmic microwave background polarization measurements~\cite{CMB-S4:2020lpa}. Despite these extensive efforts, the ultra-high-frequency regime well above the $\mathrm{kHz}$ band remains largely unexplored~\cite{Aggarwal:2025noe}.

Recently, this high-frequency frontier has attracted renewed theoretical and experimental interest. Because no known standard astrophysical objects are sufficiently small and dense to radiate efficiently at frequencies $\gtrsim 10\ \mathrm{kHz}$, high-frequency gravitational waves (HFGWs) offer a uniquely clean background, free of conventional astrophysical noise~\cite{Aggarwal:2025noe}. Candidate sources in this regime span a wide array of exotic, beyond-Standard-Model phenomena, including light primordial black holes~\cite{Franciolini:2022htd, Dolgov:2011cq, Ireland:2023avg}, boson stars~\cite{Liebling:2012fv, Visinelli:2021uve}, superradiant boson clouds~\cite{Arvanitaki:2014wva, Brito:2015oca, Brito:2017zvb}, and early-universe first-order phase transitions~\cite{Kamionkowski:1993fg, Espinosa:2010hh, Hindmarsh:2013xza, Hindmarsh:2017gnf, Kersten:2024ucm, Biswas:2025rzs} (see Ref.~\cite{Roshan:2024qnv} for a recent review).

To probe these candidates, specialized detection methodologies have been proposed, spanning advanced laser interferometry~\cite{Akutsu:2008qv, Ackley:2020atn}, optically levitated sensors~\cite{Aggarwal:2020umq}, mechanical resonators~\cite{Goryachev:2014yra, Goryachev:2014nna, Gottardi:2007zn}, and superconducting rings~\cite{Anandan:1982is}. In this work, we focus on resonant microwave cavities, which exploit a strong background magnetic field to transduce GWs into observable photons via the inverse Gertsenshtein effect~\cite{gertsenshtein1962, zeldovich1974emgw, Berlin:2021txa}. Our framework targets this phenomenon by deploying superconducting qubits as localized quantum sensors to extract the ultra-weak induced electromagnetic signal.

While this cavity architecture shares core physical principles with standard axion haloscopes~\cite{CAPP:2020utb, CAPP:2024dtx, Asztalos2010, ADMX:2018gho, ADMX:2021nhd}, allowing existing facilities to be repurposed for HFGW searches~\cite{Kim:2025izt,Domcke:2024eti}, conventional setups are fundamentally limited by their reliance on macroscopic excess-power measurements. These axion detectors are optimized for a spin-0 pseudoscalar field, which induces an azimuthally symmetric current perfectly coupled to the $m=0$ angular mode (typically the $\mathrm{TM}_{010}$ resonance). In contrast, the spin-2 tensor nature of a co-aligned GW enforces a characteristic, quadrupolar $|m|=2$ spatial signature~\cite{Berlin:2021txa}. By exploiting this distinct spatial profile, we can strategically position localized quantum sensors to bypass global cavity-power limits and enable a targeted readout of the GW-induced field.

To this end, we introduce a novel quantum-sensing paradigm for HFGW detection integrating three core mechanisms:
\begin{itemize}[leftmargin=*]
\item \textbf{Inverse Gertsenshtein transduction}~\cite{gertsenshtein1962, Boccaletti:1970pxw, zeldovich1974emgw}: A GW traversing a static magnetic field induces a coherently oscillating electromagnetic current.
\item \textbf{Direct quantum excitation}~\cite{Chen:2022quj, Chen:2024aya}: Bypassing classical macroscopic amplification, this induced field directly excites localized superconducting transmon qubits positioned at the cavity's electric-field hot spots (e.g., the $\mathrm{TE}_{212}$ mode).
\item \textbf{Collective multiqubit enhancement}~\cite{Chen:2023swh, Dong:2025mdk}: Configuring these local sensors as an entangled quantum register utilizing symmetric Dicke states provides a profound matrix-element enhancement over independent-qubit counting, translating into a fundamental scaling advantage.
\end{itemize}

We evaluate these protocols against the conventional excess-power readout. 
As we will demonstrate, replacing the macroscopic power measurement with an 800-qubit global Dicke register achieves a strain sensitivity scaling of $h_{\min}^{\rm Dicke}\propto n_q^{-3/4}$, suppressing the standard cavity-power benchmark by up to five orders of magnitude. 
When benchmarked against the representative geometries and fields of existing haloscopes, this collective quantum advantage decisively mitigates the profound coupling hierarchy between the axion interaction ($g_{a\gamma\gamma}$) and the Planck-scale-suppressed gravitational interaction ($1/M_{\rm Pl}$). 
These idealized upper-envelope projections set a concrete target for future open-quantum-system designs, demonstrating the transformative potential of collective quantum sensing for HFGW searches.

The paper is organized as follows. In Sec.~\ref{sec:Theoretical_Setup}, we derive the HFGW-induced electromagnetic current in the proper-detector (PD) frame. In Sec.~\ref{sec:Electromagnetic_Signal}, we solve for the resulting cavity response, extracting the $|m|=2$ selection rule and establishing the quadrupolar $\mathrm{TE}_{212}$ mode as our design target. In Sec.~\ref{sec:Signal_Detection}, we contrast conventional cavity-power measurements with our proposed independent-qubit and Dicke-state readouts, evaluating their respective sensitivity scalings. In Sec.~\ref{sec:axion_DM}, we apply this framework to the representative parameters of existing axion-haloscope cavities. Finally, we discuss the implications, open-system hardware challenges, and future prospects in Sec.~\ref{sec:Conclusion}.

\section{Gravitational-wave induced electromagnetic current}
\label{sec:Theoretical_Setup}

In the presence of a gravitational wave (GW), the spacetime metric is perturbed as
\begin{align}\label{eqn:metric}
    g_{\mu \nu} = \eta_{\mu \nu} + h_{\mu \nu}, \qquad g^{\mu \nu} = \eta^{\mu \nu} - h^{\mu \nu} \, ,
\end{align}
where $\eta_{\mu \nu} = \mathrm{diag}(+, -, -, -)$ is the flat Minkowski background metric, and $h_{\mu \nu} \ll 1$ represents a small gravitational perturbation. The interaction between the GW and the electromagnetic (EM) field is governed by the action
\begin{align}\label{eqn:ME_interaction}
    S_{\text{int}} 
    &= -\int d^4 x \, \frac{1}{2} h_{\mu \nu} T^{\mu \nu}_{(\mathrm{EM})} \nonumber \\
    &= -\int d^4 x \, \frac{1}{M_{\rm Pl}} \tilde{h}_{\mu \nu} T^{\mu \nu}_{(\mathrm{EM})},
\end{align}
where the gravitational perturbation couples to the energy-momentum tensor of the EM field, which in a flat background is given by $T_{\mathrm{EM}}^{\mu \nu} = - F^{\mu\alpha} F^{\nu}{}_{\alpha} + \frac{1}{4}\eta^{\mu\nu} F_{\alpha\beta}F^{\alpha\beta}$. 
The second equality reformulates the interaction using the canonically normalized gravitational field $\tilde{h}_{\mu\nu} \equiv \frac{M_{\rm Pl}}{2} h_{\mu \nu}$ (carrying mass dimension $[\tilde{h}_{\mu\nu}] = M^1$), where $M_{\rm Pl}=1/\sqrt{8\pi G}$ is the reduced Planck mass.

Expanding the covariant Maxwell equations to linear order in the metric perturbation yields the effective four-current:
\begin{align}\label{eqn:eff_curr}
    j_{\mathrm{eff}}^{\mu} \equiv \partial_{\nu} \left( \frac{1}{2} h F^{\mu\nu} + h^{\nu}{}_{\alpha} F^{\alpha\mu} - h^{\mu}{}_{\alpha} F^{\alpha\nu} \right),
\end{align}
where $h \equiv \eta^{\mu\nu}h_{\mu\nu}$ denotes the trace of the metric perturbation. 
This effective current acts as a source for new EM waves, thereby transducing gravitational energy into a detectable EM signal.
\footnote{For comparison, the analogous effective current sourced by an axion-EM interaction is $j_{\mathrm{Axion}}^\mu = \frac{1}{2} g_{a \gamma \gamma} (\partial_\nu a) \epsilon^{\mu \nu \rho \sigma} \bar{F}_{\rho \sigma}$.}

It is important to note that $j_{\mathrm{eff}}^{\mu}$ is not a true covariant four-current.
Because the linearized metric perturbation $h_{\mu\nu}$ is gauge-dependent under infinitesimal coordinate transformations, the effective current is not invariant under a change of frame at $\mathcal{O}(h)$. 
Nevertheless, once a specific coordinate frame is fixed, $j_{\mathrm{eff}}^{\mu}$ enters the flat-space Maxwell equations exactly as an ordinary electromagnetic source current.
It thus dictates the generation of the signal EM field when a GW permeates a background EM field~\cite{gertsenshtein1962, Boccaletti:1970pxw, zeldovich1974emgw}. 
In this work, we focus exclusively on this direct GW--EM coupling, neglecting indirect EM signals that could arise from acoustic or mechanical deformations of the cavity walls~\cite{fortini1982fermi}.

To explicitly evaluate the effective current in Eq.~\eqref{eqn:eff_curr}, we adopt the proper-detector (PD) frame~\cite{Marzlin:1994ia, Rakhmanov:2014noa, Maggiore:2007ulw, Berlin:2021txa} (detailed in Appendix~\ref{app:PD_frame}). 
In this frame, the metric is constructed via a multipole expansion of the Riemann tensor centered on the detector.
Because the linearized Riemann tensor $R_{\mu\nu\rho\sigma}$ is gauge-invariant at leading order in the strain~\cite{Maggiore:2007ulw}, we can evaluate it using the standard transverse-traceless (TT) gauge and directly substitute the result into the PD-frame expansion. 
For a monochromatic GW propagating along the $+\hat{z}$ direction, the TT-gauge metric perturbation takes the form
\begin{align}\label{eqn:h_TT}
    h_{ij}^{\mathrm{TT}}(t,z) =
    \begin{pmatrix}
    h_{+} & h_{\times} & 0 \\
    h_{\times} & -h_{+} & 0 \\
    0 & 0 & 0
    \end{pmatrix}
    e^{i\omega_g (t - z)},
\end{align}
where $h_+$ and $h_{\times}$ denote the plus and cross polarization amplitudes~\cite{Berlin:2021txa}. 
Substituting this expression into the PD-frame expansion yields the metric components exactly for an arbitrary GW wavelength. 
The explicit expressions, provided in Appendix~\ref{app:PD_frame}, remain valid even in the resonant regime where $\omega_{g}L_{\mathrm{det}} \sim \mathcal{O}(1)$. Because the GW interaction is parametrized entirely by the TT-gauge amplitudes $h_+$ and $h_{\times}$, all projected sensitivities in this work are expressed in terms of the strain that would be measured by a distant observer.
Eqs.~\eqref{eqn:eff_curr} and \eqref{eqn:h_TT}  uniquely determine the effective current once the GW strain $h_{+,\times}$ and frequency $\omega_g$ are specified.

Because a resonant cavity responds most strongly to a coherent signal whose frequency coincides with its resonance, the proposed experiment is therefore most sensitive to monochromatic HFGW.
As a well-motivated benchmark source in the MHz--GHz region, we consider graviton emission from a superradiant axion cloud around a rotating black hole~\cite{Arvanitaki:2014wva,Brito:2015oca, Aggarwal:2025noe}, following the approach of Ref.~\cite{Kim:2025izt}.
In this scenario, an axion pair of mass $m_a$ annihilates into a single graviton, yielding an emission frequency of
\begin{equation}\label{eqn:fGW_superradiance}
    f_{\mathrm{GW}} = \frac{m_a}{\pi} \approx 5.3~\mathrm{GHz} \left(\frac{m_a}{11~\mu\mathrm{eV}}\right).
\end{equation}
The black hole mass is set by the typical parameter $\alpha = G M_{\rm BH} m_a$, which remains below unity in the non-relativistic regime.
For the non-relativistic condition $\alpha < 1$, choosing $\alpha = 0.1$ then gives~\cite{Aggarwal:2025noe, Arvanitaki:2014wva, Kim:2025izt}
\begin{align}
    M_{\rm BH} = 1.2 \times 10^{-6} M_{\odot} \left( \frac{\alpha}{0.1} \right) \left( \frac{11~\mu\mathrm{eV}}{m_a}\right).
\end{align}
The corresponding expected GW strain is~\cite{Aggarwal:2025noe, Kim:2025izt}
\begin{align}
    h_0 \approx 1.0 \times 10^{-22} \left( \frac{\alpha}{0.1} \right)^7 \left( \frac{1 {\rm AU}}{D} \right) \left( \frac{M_{\rm BH}}{10^{-6} M_\odot} \right),
\end{align}
where $D$ is the distance to the source in astronomical units.
This emission does not last forever. 
The cloud discharges through annihilation over a finite coherence time~\cite{Aggarwal:2025noe, Kim:2025izt}
\begin{align}\label{eqn:tau_s_superradiance}
    \tau_s \approx 4.7~\mathrm{days}\left(\frac{\alpha}{0.1}\right)^{-15}\left(\frac{M_{\rm BH}}{10^{-6}M_\odot}\right).
\end{align}
We fix $\alpha=0.1$ throughout, since at this value the cloud radiates for several days, long enough to stay monochromatic over the interrogation times considered below.
The choice is a delicate one: $\tau_s$ falls off as the fifteenth power of $\alpha$, so pushing $\alpha$ only slightly higher, past roughly 0.12--0.13, cuts the emission time below a day.

Since this work aims to determine the sensitivity to the dimensionless strain, $h_0$, for a monochromatic HFGW lasting at least the data-taking period, we do not examine the origin of the signal and let the reader translate our bounds to their preferred model.

\section{Electromagnetic response in the cavity}
\label{sec:Electromagnetic_Signal}

Having fixed the GW-induced source current in Sec.~\ref{sec:Theoretical_Setup}, we now solve the cavity Maxwell equations that it drives. 
We expand the induced field in cavity eigenmodes, establish the co-aligned $|m|=2$ azimuthal selection rule, and use $\rm TE_{212}$ as a representative mode for locating local-field hot spots.
The selection rule alone does not establish that $\rm TE_{212}$ is globally optimal among all $\rm TE_{2nl}$ modes.

\begin{figure}[t]
    \centering
    \begin{subfigure}{0.49\columnwidth}
        \centering
        \includegraphics[width=\linewidth]{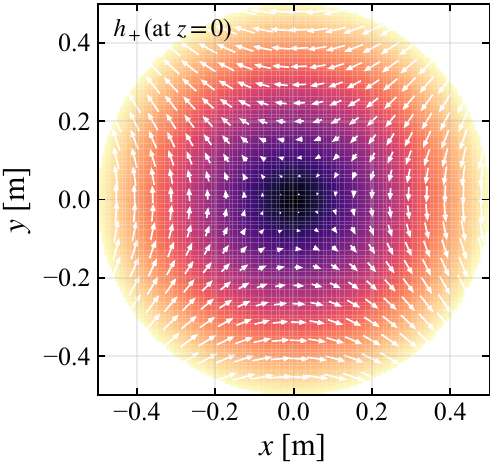}
        \caption{\(h_+\) polarization.}
        \label{fig:Plus_Current}
    \end{subfigure}
    \hfill
    \begin{subfigure}{0.49\columnwidth}
        \centering
        \includegraphics[width=\linewidth]{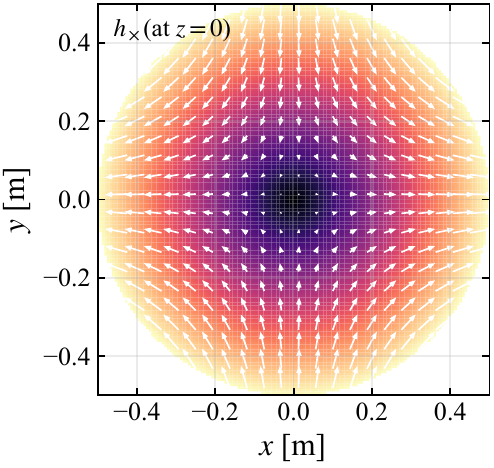}
        \caption{\(h_\times\) polarization.}
        \label{fig:Cross_Current}
    \end{subfigure}
    \caption{
    GW-induced effective current in the transverse $xy$ plane at the center of the cavity, for the $h_+$ (a) and $h_\times$ (b) polarizations.
    The color scale is the normalized magnitude and the arrows the transverse direction.
    }
    \label{fig:Effective_Current}
\end{figure}

\subsection{Mode expansion}
\label{sec:Mode_Expansion}

As explained in the previous Section, we can calculate the effective current $j^\mu_{\rm eff} = (\rho_{\rm eff}, {\bf j}_{\rm eff})$ in Eq.~\eqref{eqn:eff_curr} as additional source terms in the inhomogeneous Maxwell equations
\begin{align}\label{eqn:Max_inhomo}
    &{\bf \nabla} \cdot {\bf E}  = \rho_{\rm eff} \\
    &{\bf \nabla} \times {\bf B} - \partial_t {\bf E} = {\bf j}_{\rm eff}
\end{align}
where we retain only the effective current, so that all fields are of order $\mathcal{O}(h)$.
On the other hand, the homogeneous  Maxwell equation (Gauss's law for magnetism $\nabla \cdot {\bf B} =0$ and Faraday's law, $\nabla \times {\bf E} + \partial_t {\bf B} = 0$) are not modified by the GW.
Indeed, this is because these Maxwell equations come from a topological equation of motion $dF = 0$ (where $d$ is the exterior derivative and $F$ is the field strength two-form) which does not involve the metric $g_{\mu \nu}$.
This will be of practical importance because the homogeneous Maxwell equations determine the resonant cavity modes, and thus will see that the change in modes induced by the tidal force of the GW affects our signal at $\mathcal{O}(h^2)$.

Combining the homogeneous and inhomogeneous Maxwell equations gives the curl–curl equation~\cite{collin1990field,jackson1999classical,hill2009electromagnetic}
\begin{align}\label{eqn:curlcurl}
    \nabla\times\left(\nabla\times\mathbf{E}\right) + \partial_t^2 \mathbf{E} = -\partial_t \mathbf{j}_{\rm eff}.
\end{align}
In this form, the effective charge density does not appear explicitly.
Therefore, inside the cavity, the induced electric field can be decomposed into the normal modes of the resonator,
\begin{align}\label{eqn:mode_expansion}
    \mathbf{E}(\mathbf{x},t)
    =
    \sum_c e_c(t)\,\mathbf{E}_c(\mathbf{x}) ,
\end{align}
where $\mathbf{E}_c(\mathbf{x})$ denotes the spatial profile of the $c$-th cavity eigenmode, and $e_c(t)$ is its time-dependent excitation amplitude.
The spatial mode function ${\bf E}_c$ satisfies the boundary conditions that
\begin{subequations}\label{eqn:mode_conditions}
\begin{align}
    &\nabla^2 \mathbf{E}_c(\mathbf{x}) = - \omega_c^2 \mathbf{E}_c(\mathbf{x}),
    \label{eqn:mode_condition_1}
    \\
    &\int_{V_{\rm Cav}} d^3 {\bf x} \ {\bf E}_{c}^* ({\bf x}) \cdot {\bf E}_{c'} ({\bf x})  =  \delta_{c c'} \int_{V_{\rm Cav}} d^3 {\bf x} \ |{\bf E}_{c'} ({\bf x})|^2
    \label{eqn:mode_condition_2}
\end{align}
\end{subequations}
where $\omega_c$ is the resonant frequency of mode $c$, and $V_{\rm Cav}$ is the volume of the cavity.
The above relations are supplemented by the boundary condition $\hat{n} \times {\bf E} = 0$ over the surface of the cavity, where $\hat{n}$ is the normal unit vector of the boundary.
Note that formally this boundary condition is unchanged even in the presence of the GW, since it follows from the form of Faraday’s law, as mentioned above (detailed in Appendix~\ref{app:mode_expansion}).

Using Eqs.~\eqref{eqn:mode_condition_1} and \eqref{eqn:mode_condition_2}, substituting the mode expansion of Eq.~\eqref{eqn:mode_expansion} yields an equation for the mode coefficients $e_c(t)$,
\begin{align}\label{eqn:Wave_eqn}
    \left( \partial_t^2 + \frac{\omega_c}{Q_c} \partial_t + \omega_c^2 \right) e_c(t) = - \frac{\int_{V_{\rm Cav}} d^3 {\bf x} \ {\bf E}_c^* ({\bf x}) \cdot \partial_t {\bf j}_{\rm eff}}{\int_{V_{\rm Cav}} d^3 {\bf x} \ |{\bf E}_c ({\bf x})|^2}
\end{align}
where the mode-dependent quality factor $Q_c$ arises from losses in the cavity.
In principle, $\omega_c$ and $V_{\rm Cav}$ appearing in the above equation are perturbed by the tidal force of the GW at $\mathcal{O}(h)$.
However, if in the absence of the GW there is no power in cavity modes, this effect corrects our signal only by terms of $\mathcal{O}(h e_c)$ and $\mathcal{O}(hj_{\rm eff})$, both of which are $\mathcal{O}(h^2)$.

Taking the GW to be a monochromatic and on resonance with the cavity mode, i.e. ${\bf j}_{\rm eff} ({\bf x}, t) = e^{i \omega_g t} {\bf J}_{\rm eff} ({\bf x})$ where $\omega_g \simeq \omega_c$, the solution to Eq.~\eqref{eqn:Wave_eqn} is enhanced by the large quality factor $Q_c \gg 1$. 
The steady-state form of the excited signal electric field ${\bf E}_{\rm sig} = e_c {\bf E}_c$ is then given by
\begin{align}\label{eqn:E_sig}
    \mathbf{E}_{\rm sig}(\mathbf{x},t)=-
    \frac{
    \int_{V_{\rm Cav}} d^3{\bf x}'\,
    \mathbf{E}_c^{*}\cdot \mathbf{J}_{\rm eff}\, 
    }{
    \int_{V_{\rm Cav}} d^3{\bf x}'\, |\mathbf{E}_c|^2
    }
    \frac{Q}{\omega_g}\,
    \mathbf{E}_c(\mathbf{x}) e^{i\omega_g t},
\end{align}
where for convenience we drop the subscript from the quality factor and assume that it is the same for each mode, $Q = Q_c$.
Note that if there are multiple degenerate modes at $\omega_g$, we must sum over all modes with their own time functions $e_c(t)$ to obtain the signal field (see Appendix~\ref{app:detune response}).

We define the dimensionless overlap factor
\begin{align}\label{eqn:overlap_factor}
    \eta_c \equiv 
    \frac{
    \int_{V_{\rm Cav}} d^3\mathbf{x} \,
    \mathbf{E}_c^*(\mathbf{x})\cdot
    \mathbf{J}_{\rm eff}(\mathbf{x}) }
    {N_c^{1/2} V_{\rm Cav}^{5/6} B_0 h_0 \omega_g^2} ,
\end{align}
where $N_c \equiv \int_{\rm Cav} d^3x\, |\mathbf{E}_c(\vec x)|^2$ and $h_0 = |h_+| \ \text{or} \ |h_\times|$ is the magnitude of GW strain.
And we can rewrite Eq.~\eqref{eqn:E_sig} in a compact form as
\begin{align}\label{eqn:E_sig_overlap}
    \mathbf E_{\rm sig}(\mathbf x,t)= - B_0 h_0 \omega_g Q V_{\rm Cav}^{5/6}\sum_{c \in{\cal D}} \frac{1}{{\sqrt{N_c}}}\eta_c \mathbf{E}_c (\mathbf{x})  e^{i\omega_g t}.
\end{align}
where $\cal D$ is the degenerate subspace.

\subsection{Cylindrical resonant cavity \& overlap factor}
\label{sec:cyl_cavity_overlap_factor}

We consider a cylindrical cavity of radius $R$ and length $L$ with perfectly conducting walls. 
Such geometries are widely employed in high-precision electromagnetic resonance experiments, including axion haloscope searches such as ADMX~\cite{Asztalos2010, ADMX:2018gho, ADMX:2021nhd}, HAYSTAC~\cite{HAYSTAC:2018rwy}, CAPP~\cite{CAPP:2020utb, CAPP:2024dtx}, QUAX~\cite{Alesini:2019ajt, QUAX:2024fut}, and ORGAN~\cite{McAllister:2017lkb}, due to their well-understood mode structure and high quality factors.

The electromagnetic eigenmodes of a cylindrical cavity are classified as transverse magnetic (TM) and transverse electric (TE) modes with respect to the cavity axis $z$, satisfying $B_z = 0$ and $E_z = 0$, respectively. 
Each mode is labeled by a set of integers $(m,n,l)$, where $m \in \mathbb{Z}$ is the azimuthal index, $n=1,2,\dots$ is the radial index, and $l$ is the longitudinal index. 
The detailed mathematical formalism for the cylindrical cavity modes, including the
mode functions, normalization, and basis conventions, is provided in  Appendix~\ref{app:mode_expansion}.

The spin-2 nature of gravitational waves imposes an azimuthal selection rule, as the cross-shaped current pattern in Figure~\ref{fig:Effective_Current} shows. 
For a co-aligned GW, static magnetic field, and cavity axis, the induced effective current carries angular momentum corresponding to helicities $m=\pm2$, so that only cavity modes with $|m|=2$ have a nonvanishing azimuthal overlap. 
We therefore adopt the $\rm TE_{212}$ mode as a representative design mode because it provides convenient off-axis local electric-field antinodes in the geometry studied here. A global mode optimization would require comparing all accessible modes together with their mode-dependent losses.

The cavity size sets the resonant frequency of the GW-induced field, $\omega_g \simeq \omega_{mnl}$, which is roughly given by
\begin{align}
    \omega_{mnl}
    \sim
    \sqrt{
    \left(\frac{x_{mn}}{R}\right)^2
    +
    \left(\frac{l\pi}{L}\right)^2
    } \, ,
    \label{eqn:mode_freq_approx}
\end{align}
where $x_{mn}$ is the Bessel-function zero for the corresponding mode.
Therefore, to detect HFGWs in the frequency range $\mathcal{O}(1)\sim\mathcal{O}(10)\,\mathrm{GHz}$, the proper cavity size should be around the microwave wavelength scale, where $V_{\rm Cav}$ is around $\mathcal{O}(10)\sim\mathcal{O}(10^5)\,\mathrm{cm}^3$.
In this size range, the cavity can be resonantly excited by the GW-induced effective current, and the induced electromagnetic signal can be enhanced by the cavity quality factor.

\subsubsection{Rotational dependence}
\label{sec:cyl_cav_rot}

\begin{figure}[t] 
    \centering
    \includegraphics[width=0.8\columnwidth]{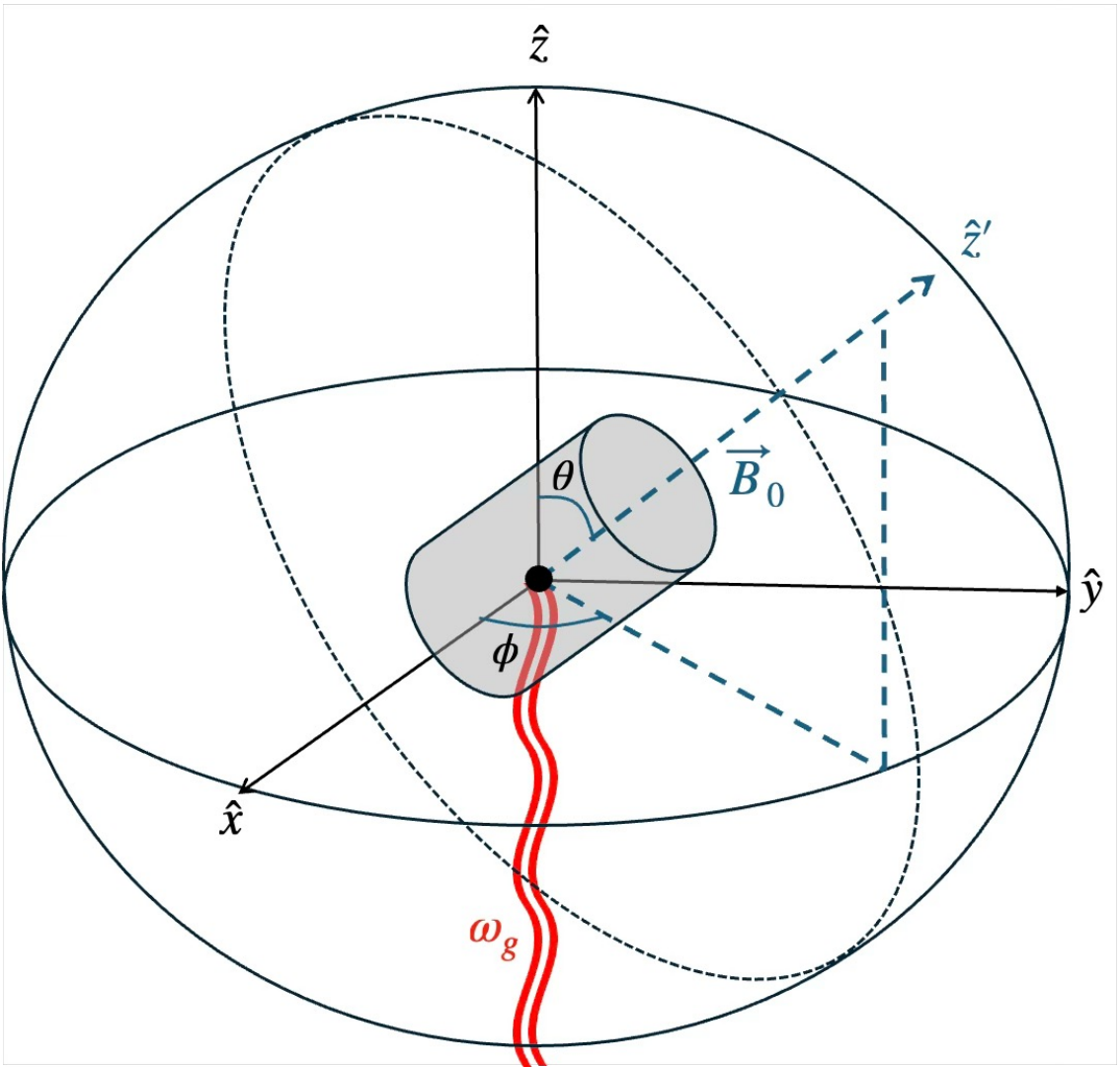} 
    \vspace*{0.3cm}
    \caption{
    Geometric configuration for an incoming GW and a cylindrical cavity.
    To describe arbitrary GW propagation, the GW incidence is fixed along the $\hat{z}$ axis with frequency $\omega_g$ assuming that $h_{\mu \nu} \propto e^{i \omega_g (t-z)}$.
    The orientation of the cavity is parameterized by $\theta$, representing the polar angle  between the GW propagation vector and the longitudinal axis of the cylinder ($\hat{z}'$), and the azimuthal angle $\phi$, defined as the angle between the $\hat{x}$ axis and the projection of the cavity axis onto the $xy$ plane. 
    The external $\vec{B}_0$ field is aligned with $\hat{z}'$-axis.
    }
    \label{fig:Cavity_Rotation} 
\end{figure}

When the GW propagation, cavity axis, and static magnetic field are co-aligned, this induced current forms a distinct quadrupolar pattern in the transverse plane. 
However, once the GW is not incident along the cavity axis, the cylindrical symmetry breaks and additional azimuthal components become accessible.
To understand and quantify the broken symmetry for an arbitrary incoming GW, we use two coordinate systems sharing the same origin at the cavity center. 
The first is the global coordinate system $(x,y,z)$, in which the incident GW is taken to propagate along the $z$-axis. 
The second is the local coordinate system $(x',y',z')$, adapted to the cylindrical cavity, with the $z'$-axis chosen along the cavity axis. 
As shown in Figure~\ref{fig:Cavity_Rotation}, the angle $\theta$ denotes the angle between the GW propagation direction and the cavity axis, while $\phi$ specifies the azimuthal orientation of the cavity about the global $z$-axis.

\begin{figure}[t] 
    \centering
    \includegraphics[width=0.97\columnwidth]{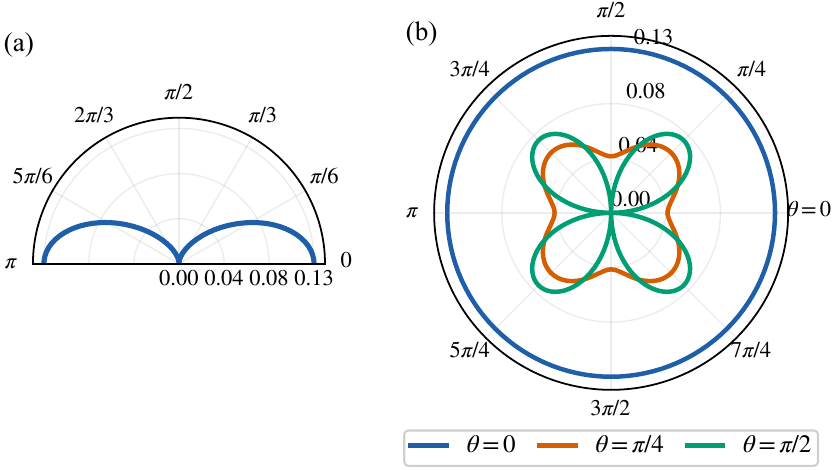} 
    \caption{
    Angular dependence of the GW--cavity overlap factor magnitude [Eq.~\ref{eqn:eta_deg}] for the $\rm TE_{212}$ mode.
    $\rm TE_{212}$ has a degenerate pair $m=\pm 2$ in basis $e^{im\phi}$ [Eqs.~\ref{eqn:TE_mode}].
    Panel (a) shows the polar-angle dependence as a function of $\theta$ at fixed azimuthal orientation ($\phi = 0$).
    Panel (b) shows the azimuthal dependence as a function of $\phi$ for representative values of $\theta=0$, $\pi/4$, and $\pi/2$.
    }
    \label{fig:Angular_Dependence} 
\end{figure}

Figure~\ref{fig:Angular_Dependence} shows the angular dependence of the basis-invariant degenerate-pair norm $|\eta_{\rm deg}|$ for the $\rm TE_{212}$ mode, rather than the coherent local field seen by a specified qubit dipole.
We fix the GW propagation direction along the $\hat z$ axis and rotate the cylindrical cavity axis $\hat z'$ by the polar and azimuthal angles $(\theta,\phi)$, as defined in Figure~\ref{fig:Cavity_Rotation}.  
Panel (a) shows the dependence on the polar angle $\theta$ for a fixed azimuthal orientation, while panel (b) shows the dependence on the azimuthal angle $\phi$ for representative values of $\theta=0$, $\pi/4$, and $\pi/2$.  
The overlap is maximal when the cavity axis is aligned or anti-aligned with the GW propagation direction and is suppressed near the transverse orientation. 
The nontrivial $\phi$ dependence for fixed $\theta$ reflects the projection of the cavity mode onto the GW polarization basis.

Away from co-alignment, the overlap factor exhibits different mode preferences as the rotational symmetry is broken.
As shown in Figures~\ref{fig:diff_mode_theta} and~\ref{fig:diff_mode_phi} of Appendix~\ref{app:rot_response}, the GW-induced effective current no longer carries a pure $m=\pm2$ azimuthal structure once the cavity axis is tilted with respect to the GW propagation direction.
Instead, the rotation mixes different azimuthal components, allowing cavity modes with $m\neq\pm2$ to acquire finite overlap.
Consequently, while the $\mathrm{TE}_{212}$ mode remains a representative high-overlap choice near the co-aligned configuration, other cavity modes can become comparable or even dominant for particular orientations.
This behavior provides additional flexibility in designing cavity geometries and operating configurations for all-sky GW searches, where the incident direction is generally unknown.

\begin{figure*}[htbp]
    \centering
    \includegraphics[width=0.95\textwidth]{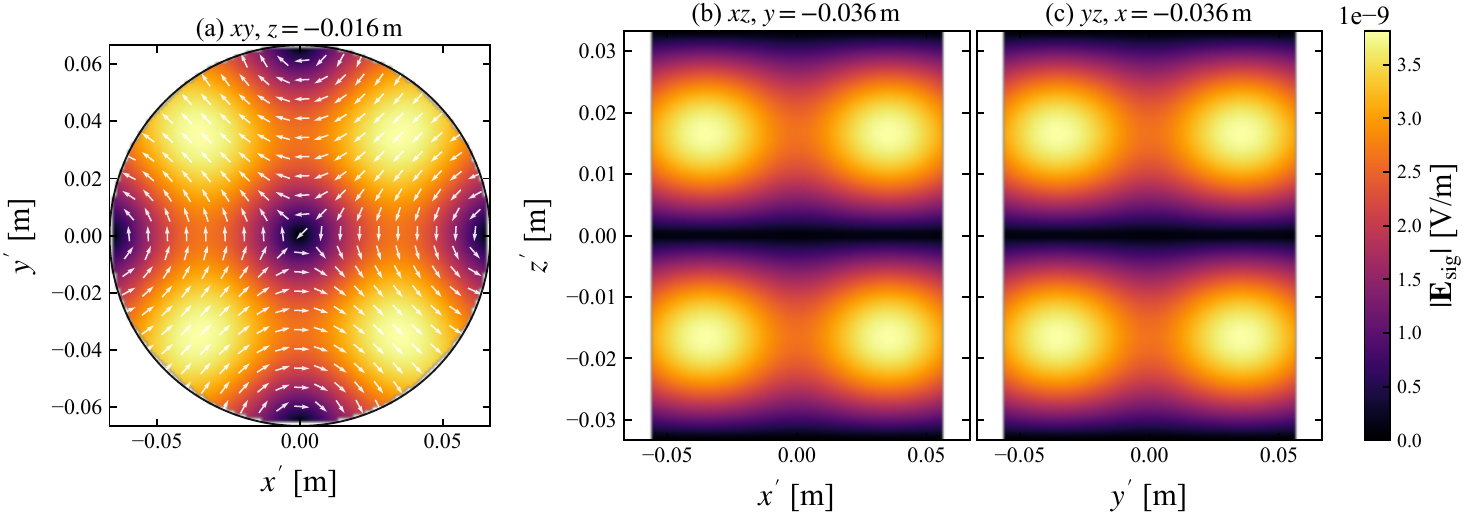}
    \caption{
    Spatial profile at hot spots of the induced signal electric field for the TE$_{212}$ mode.
    Panels (a), (b), and (c) show representative $xy$, $xz$, and $yz$ slices through the cylindrical cavity, respectively.
    The benchmark parameters are $R=6.67~\mathrm{cm}$, $L=6.67~\mathrm{cm}$ (so that $f_{\mathrm{TE}_{212}}=5~\mathrm{GHz}$), $B_0=5.0~\mathrm{T}$, $Q=10^{5}$, and $h_0=10^{-23}$, for a co-aligned $h_+$ polarized GW on the $\mathrm{TE}_{212}$ resonance; the plotted field is the steady-state coherent amplitude of Eq.~\eqref{eqn:E_sig_overlap} with the mode normalization of Eq.~\eqref{eqn:mode_condition_2} (peak $\bar{\mathbf{E}}_{\rm sig}=3.8\times10^{-9}~\mathrm{V/m}$).
    The color scale denotes the magnitude $\bar{\mathbf{E}}_{\rm sig}$ in V/m, while the arrows in panel (a) indicate the transverse field direction in the $xy$ plane.
    }
    \label{fig:TE212_SignalField}
\end{figure*}

\begin{figure}[htbp] 
    \centering
    \includegraphics[width=0.75\columnwidth]{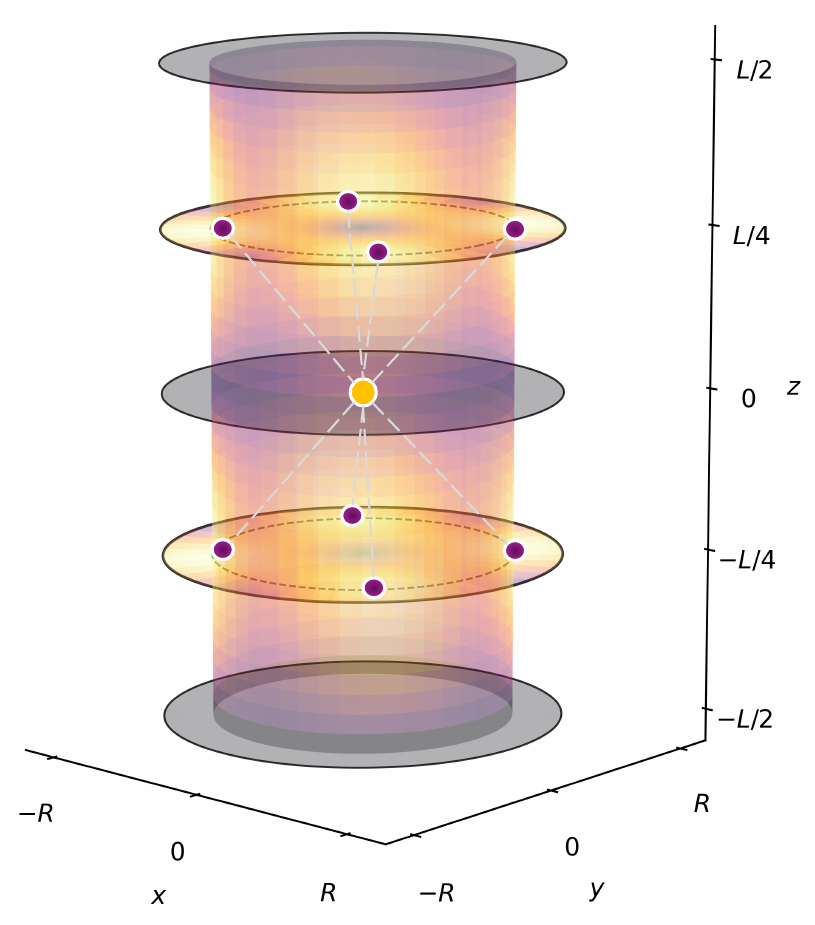} 
    \caption{
    Tesseract-like qubit structure for qubit readout.
    Sensor qubits (purple) sit at the off-axis $\mathrm{TE}_{212}$ hot spots on the $z=\pm L/4$ antinodal planes. The central ancilla qubit (yellow) sits at the center of the cavity, where it is a field node for the $\rm TE_{212}$ mode and collects the accumulated signal through the readout circuits of Figure~\ref{fig:Dicke_circuit}.
    }
    \label{fig:Cavity_Qubit} 
\end{figure}

\subsubsection{Electric field signal}
\label{sec:cyl_E_sig}

The polarization dependence of the GW-induced effective current is illustrated in Figure~\ref{fig:Effective_Current}.  
The two GW polarizations generate distinct transverse current patterns in the cavity cross section, which in turn lead to different projections onto the TE$_{212}$ cavity mode.

Figure~\ref{fig:TE212_SignalField} shows the spatial structure of the induced signal electric field projected onto the $\mathrm{TE}_{212}$ cavity mode. 
The transverse $xy$ slice exhibits the in-plane field pattern together with the relative field direction, while the $xz$ and $yz$ slices show the longitudinal nodal structure of the mode. 
The result illustrates that the GW-induced response is localized according to the cavity-mode profile rather than being spatially uniform throughout the cavity volume.

This localized structure is what the qubit readout exploits.
The transverse field is largest on the two antinodal planes at $z=\pm L/4$ and vanishes on the midplane, so a transmon placed at an off-axis hot spot samples the field where it is strongest, while the cavity axis remains a node well-suited to a non-sensing ancilla (Figure~\ref{fig:Cavity_Qubit}).
The same spatial variation distinguishes the two readout strategies considered in this work. The conventional cavity readout measures the total electromagnetic energy stored in the resonator at the mode frequency, whereas the qubit detector probes the local field at a fixed position.
In the following, we restrict our analysis to the $h_+$ polarization, with the cavity axis co-aligned with the principal polarization frame of the incoming gravitational wave.
The $h_\times$ polarization follows from the same construction and is equivalent to a $\pi/4$ rotation in the transverse plane relative to the $h_+$ polarization. 
Consequently, for the cylindrical cavity considered here, it is already accounted for by the azimuthal-angle average.
Now, we turn to these readouts in Sec.~\ref{sec:Signal_Detection}.

\section{Signal Detection \& Sensitivity}
\label{sec:Signal_Detection}

The GW-induced effective current [Eq.~\ref{eqn:eff_curr}] excites an electromagnetic mode in the cavity and produces the signal electric field given in Eq.~\eqref{eqn:E_sig_overlap}. 
In this section, we discuss two possible readout protocols.
The first is the conventional cavity-power readout, which measures the power stored and dissipated in the resonantly excited cavity mode~\cite{Berlin:2021txa, Kim:2025izt}. 
The second is the qubit readout, where a qubit is placed near the maximum of the signal electric field and detects the local field through its excitation probability~\cite{Chen:2022quj, Chen:2024aya}. 

\subsection{Cavity power readout}
\label{sec:power_readout}

In the power-readout protocol, the signal is obtained from the energy stored in the induced electric field and the cavity decay rate. 
Using the signal field in Eq.~\eqref{eqn:E_sig_overlap}, the cavity signal power is~\cite{Berlin:2021txa}
\begin{align}
    P_{\rm sig}
    & =
    \frac{\omega_c}{2Q}
    \int_{V_{\rm Cav}} d^3{\bf x}\,
    \left|{\bf E}_{\rm sig}({\bf x})\right|^2
    \label{eq:Psig_definition} \\
    & = 
    \frac{1}{2}
    Q\omega_g^3 V_{\rm Cav}^{5/3}
    B_0^2 h_0^2
    \sum_{c\in{\cal D}} |\eta_c|^2 \, ,
    \label{eq:Psig_expand}
\end{align}
where $\omega_c$ is the resonant frequency of the cavity mode $c$.
In the resonant sensitivity estimate we set $\omega_c\simeq\omega_g$.

The sensitivity is estimated by comparing the signal power with the noise power in the measurement bandwidth. 
We use the radiometer estimate for narrow-band microwave-cavity searches
\begin{align}
    \sigma_{\rm cav}
    =
    \frac{P_{\rm sig}}{K_{\rm noise}}
    \sqrt{\frac{t_{\rm int}}{\Delta f}},
    \label{eq:cavity_snr}
\end{align}
where $t_{\rm int}$ is the total integration time and $\Delta f$ is the effective bandwidth.
The noise energy per unit bandwidth is written as
\begin{align}
    K_{\rm noise}
    & =
    T_{\rm sys} \nonumber \\
    & = \omega_c \left[
    n_{\rm th}(\omega_c,T_{\rm cav})+n_{\rm add}
    \right],
    \label{eq:cavity_noise_energy}
\end{align}
where
\begin{align}
    n_{\rm th}(\omega,T)
    =
    \frac{1}{\exp\left(\omega/ T\right)-1}
    \label{eqn:thermal_occupation}
\end{align}
is the thermal photon occupation number and $n_{\rm add} \geq \tfrac{1}{2}$ denotes the added readout noise in units of photon number including the vacuum half quantum~\cite{Jeong:2021dij}.
For the cavity power readout we set it to $\frac{1}{2}$.
For the following result, we take the effective bandwidth to be
\begin{align}
    \Delta f
    =
    \max\left(
    \frac{f_c}{Q},
    \frac{1}{t_{\rm int}},
    \frac{1}{\tau_{\rm GW}}
    \right),
    \label{eq:cavity_bandwidth}
\end{align}
which accounts for the finite cavity linewidth and the finite integration time~\cite{Berlin:2021txa}, and defines our classical comparator: an incoherent excess-power (radiometer) analysis with bandwidth bounded by the cavity linewidth.
For a sufficiently coherent source with a known or searched waveform, a phase-sensitive or matched-filter readout of the same cavity could instead integrate coherently over a bandwidth set by the signal coherence, frequency drift, and Fourier resolution, $\Delta f\sim\max(1/\tau_{\rm GW},1/t_{\rm int},\Delta f_{\rm drift})$, in which regime the classical SNR scales linearly with $t_{\rm int}$.
The comparisons below are therefore benchmarks against the standard excess-power analysis, not claims of surpassing an optimized coherent classical measurement chain; the radiometer baseline uses $\Delta f=\max(f_c/Q,1/t_{\rm int})$ while the qubit protocols assume full signal coherence over $\tau_{\rm eff}$.
Then, the corresponding parametric strain scaling is
\begin{align}
    {h_{\min}^{\rm cav}}
    & \propto \frac{\sigma^{1/2}_* K^{1/2}_{\rm noise}}{\eta B_0 Q^{1/2} \omega_g^{3/2} V^{5/6}_{\rm Cav} }\left(\frac{\Delta f}{t_{\rm int}}\right)^{1/4} \nonumber \\
    & \propto
    \frac{\sigma_*^{1/2} K_{\rm noise}^{1/2} \omega_g}{  \eta B_0 Q^{1/2}}
    \left(\frac{\Delta f}{t_{\rm int}}\right)^{1/4} \, ,
    \label{eqn:hmin_cav_counting}
\end{align}
where the cavity volume $V_{\rm cav}$ is scaled to $\omega^{-3}$ for the second line and $\sigma_*=5$ is chosen as the threshold of discovery.
Thus all quoted sensitivities correspond to the strain required to achieve a nominal $5\sigma$ detection significance.

\subsection{Independent-qubit and entangled-qubit readout}
\label{sec:qubit_readout}

\begin{figure}[!t] 
    \centering
    \includegraphics[width=0.95\columnwidth]{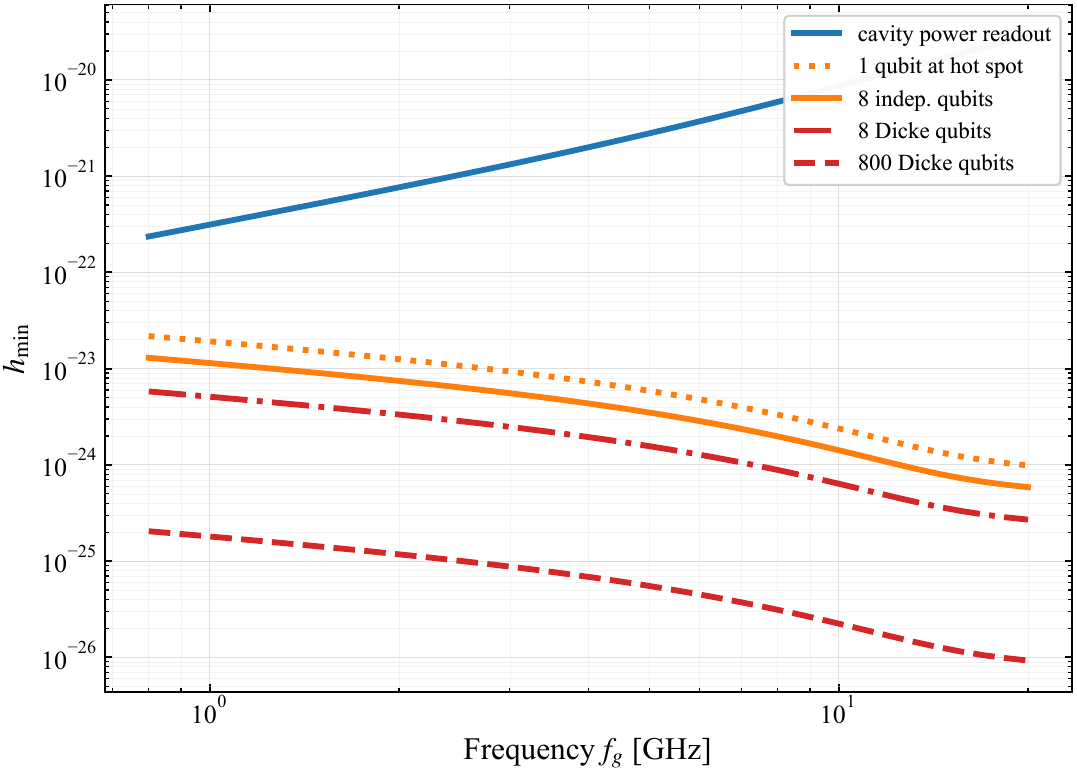} 
    \caption{
    $5\sigma$-benchmark projected strain sensitivity $h_{\min}$ for an $L=R$ cavity rescaled at each point so that the $\mathrm{TE}_{212}$ resonance tracks the GW frequency $f_g$ (with $\theta = 0$).
    The curves show the cavity-power readout (blue), one qubit at the $\mathrm{TE}_{212}$ hot spot (orange dotted), and eight qubits read out independently (orange solid), in a symmetric Dicke state for 8 qubits (red dash-dotted) and 800 qubits (red dashed). 
    The Dicke curves assume a single global Dicke register ($n_D=n_q^2/2+n_q$); the eight-site local implementation of Appendix~\ref{app:Ent_Enh} would degrade the 800-qubit strain reach by a factor of $\simeq2.8$.
    The numerical simulation uses $B_0=5\,\mathrm{T}$, $Q=10^5$, $\tau_{\rm eff}\simeq 100 {\rm \mu s}$, $C=0.1\,\mathrm{pF}$, $d=100\,\mu\mathrm{m}$, $t_{\rm int}=1\,\mathrm{min}$ per point and $p_{\rm ro} = p_{\rm op} = 0.1\%$.
    All qubit curves are computed in the prescribed-field (no-backaction) treatment of the qubit drive
    }
    \label{fig:Sensitivity_LR} 
\end{figure}

We now consider a local qubit-readout protocol of the GW-induced cavity field. 
For a qubit at $\mathbf{x}_q$ with capacitive direction $\hat{\mathbf n}_{\rm cap}$  and the specific mode $c \in {\cal D} = \{ {\rm TE}_{+ 212}, {\rm TE}_{- 212} \}$, we define the dimensionless local coherent response of the degenerate subspace by
\begin{align}
    {\cal R}_q(\mathbf{x}_q,\hat{\mathbf n}_{\rm cap})
    \equiv
    \left|
    \sum_{c\in\mathcal D}
    \eta_c
    \sqrt{\frac{V_{\rm Cav}}{N_c}}\,
    \mathbf E_c(\mathbf{x}_q)\cdot\hat{\mathbf n}_{\rm cap}
    \right|.
\end{align}
The local signal field magnitude is then
\begin{align}
    \bar E_{\rm sig}
    & = |{\bf E}_{\rm sig}| \nonumber \\
    & = Q\omega_g V_{\rm Cav}^{1/3} B_0 h_0\,
   {\cal R}_q .
    \label{eq:Ebar_qubit}
\end{align}
For the idealized peak-field benchmark used in Figure~\ref{fig:Sensitivity_LR}, we maximize this response over the qubit position and dipole orientation, $\mathcal R_q^{\rm peak}=\max_{\mathbf{x}_q,\hat{\mathbf n}_{\rm cap}} {\cal R}_q(\mathbf{x}_q,\hat{\mathbf n}_{\rm cap})$.

The effective field, Eq.~\eqref{eq:Ebar_qubit}, modifies the Hamiltonian of the transmon qubit by adding interaction terms
\begin{align}\label{eqn:q_int_H}
    H_{\rm int} & = Cd \hat{V} \mathrm{Re}\left[\bar{E}_{\rm sig} e^{i\omega_g t}\right] \nonumber \\
    & = 2 \delta \cos\left(\omega_g t - \alpha \right)(\hat{a} + \hat{a}^\dagger) \, ,
\end{align}
where $\alpha$ is the drive phase, $C$ is the capacitance, $\hat{V}$ is the voltage difference across the Josephson junction, and $d$ is the effective distance between the two plates~\cite{Chen:2024aya}.
The $\hat{a}$ and $\hat{a}^\dagger$ note the ladder operators associated with the transmon qubit ground state $|g\rangle$ and excited state $|e\rangle$, detailed in Appendix~\ref{app:Int_Ham}.
The corresponding qubit system Hamiltonian approximately has $H_0 \simeq \omega_q |e\rangle\langle e|$ with the energy ground state $|g\rangle$ taken to be zero.

Therefore, the local field excites the qubit through the electric-dipole coupling to the transmon capacitor. 
Following the direct-excitation estimate for a transmon qubit, the effective
single-qubit drive strength is
\begin{align}
    \delta
    =
    \frac{\sqrt{\omega_q C}}{2\sqrt{2}}\,
    d\,\bar{E}_{\rm sig},
    \label{eq:qubit_drive}
\end{align}
where $C$ is the qubit capacitance, $d$ is the effective capacitor separation, and $\omega_q$ is the qubit transition frequency~\cite{Chen:2022quj, Chen:2024aya}.

We adopt a prescribed-field, no-backaction benchmark in which the GW-induced electric field is first computed for the bare cavity and is subsequently applied as an externally specified classical drive to an idealized local qubit sensor.
Following the direct-excitation treatment of Refs.~\cite{Chen:2022quj,Chen:2024aya}, we tune the qubit into resonance with the signal, $\omega_q\simeq\omega_g$, and use the resonant weak-drive transition probability. 
The cavity enhances the GW-induced field through its multimode response, while the qubit samples the resulting local electric field at its position.

The same transmon dipole matrix element fixes both the vacuum cavity-qubit coupling and the coherent signal drive. 
The coupling comes from projecting that dipole onto the cavity's zero-point field at the qubit's location, and the drive strength scales with that coupling times the coherent cavity amplitude the GW generates.
For the benchmark $Q=10^5$, the cavity linewidth is $\kappa=\omega_c/Q\sim\mathcal{O}(10){\rm kHz}$, whereas the cavity--qubit coupling $\lambda_c$ is typically $\lambda_c\sim\mathcal{O}(1){\rm MHz}$~\cite{Chen:2024aya}. 
At simultaneous cavity, qubit, and signal resonance, $\omega_c\simeq\omega_q\simeq\omega_g$, the system therefore lies in the resolved strong-coupling regime, where the one-excitation eigenstates are qubit--cavity polaritons rather than independent bare excitations. 
The signal response is then distributed between the polariton branches according to their cavity and qubit admixtures, and the bare-qubit transition probability used below is not self-consistent.

Detuning the qubit by $|\omega_q-\omega_c|\gtrsim\lambda_c$ suppresses the hybridization while leaving the cavity resonant with the GW, but it simultaneously detunes the qubit from the signal and reduces its direct-excitation probability. 
We therefore present the prescribed-field result only as an idealized no-backaction upper limited benchmark. 
Appendix~\ref{app:cavity_qubit_hybridization} gives the hybrid-eigenmode and open-system formulation required for a self-consistent treatment, while the corresponding quantitative polariton-readout sensitivity curves including open system noise are left for future work.

The qubit reaches quoted below should therefore be interpreted as idealized benchmark targets. 
Unless otherwise stated, the frequency-scan benchmarks assume $B_0=5~\mathrm{T}$, $Q=10^5$, an effective qubit interrogation time $\tau_q=100~\mu\mathrm{s}$, and $T_{\rm sys}=0.1~\mathrm{K}$, together with qubit tunability across the scanned band and optimal placement and dipole alignment at the relevant electric-field hot spots.
We further assume that the qubit chips, control wiring, and readout hardware do not appreciably modify the cavity quality factor or mode profile. 
The global Dicke curves additionally assume long-range control connectivity and collective parity readout across the
full register; the eight-site construction of Appendix~\ref{app:Ent_Enh} is a more modular alternative.
Achieving these conditions in a multi-tesla magnet will require high-field-compatible qubits or field-free/compensated qubit modules, low-loading wiring, and scalable control infrastructure, which remain future experimental requirements rather than demonstrated capabilities.

Following the individual measurement protocol of Ref.~\cite{Chen:2024aya}, the qubits are initialized in $|g\rangle$ and evolved for a coherent interaction time
\begin{align}
    \tau_{\rm eff}=\min(\tau_q,\tau_{\rm GW}),
\end{align}
where $\tau_{\rm GW}$ denotes the coherence time of the GW signal and $\tau_q$ the effective coherence time of the complete qubit interrogation and readout protocol.
The latter is treated as a phenomenological benchmark parameter that captures the decoherence mechanisms relevant to the protocol, rather than as a specific device parameter. 
In the prescribed-field benchmark considered here, the GW continuously drives the cavity, and the steady-state signal field is assumed to have been established before each qubit interrogation.
The bare-cavity energy ring-down time $Q/\omega_c$ describes how quickly the energy stored in the cavity decays after the drive is removed.
Although it also sets the cavity’s transient response to changes in the drive, it does not limit the phase coherence of the steady-state field maintained by the continuous GW drive under our assumptions. 
We therefore do not include it in $\tau_{\rm eff}$.
As a result, we use $\tau_q \simeq 100~\mu{\rm s}$ for $\tau_{\rm eff}$, where the coherence is governed only by the coherence time of the qubit.

Then, in the resonance limit, the perturbative excitation probability is
\begin{align}
    p_{\rm sig}
    \simeq
    \delta^2 \tau_{\rm eff}^2,
    \qquad
    \delta\tau_{\rm eff} \ll 1 .
    \label{eq:pqubit_sig}
\end{align}
For $n_q$ independent qubits and $N_{\rm shot} = t_{\rm int}/\tau_{\rm eff}$ repetitions, the expected signal count is
\begin{align}
    N_{\rm sig}^{\rm q-ind.} = n_q \,  N_{\rm shot}\,  p_{\rm sig}
    \label{eqn:q-ind_sig} \, .
\end{align}
The signal is obtained for by repeating the qubit measurement many times. 

We consider the collective Dicke-state protocol, following Ref.~\cite{Dong:2025mdk}.
A Greenberger-Horne-Zeilinger (GHZ) register has the quadratic scaling with $n_q$ in the ideal noiseless limit~\cite{Chen:2023swh, Chen:2024aya}.
However, this enhancement relies on global phase coherence across the register, which is fragile under decoherence.
Under amplitude damping the GHZ signal transition becomes indistinguishable from relaxation, and the collective advantage collapses.
For an optimal excitation-number-resolving measurement, the symmetric Dicke state has been shown to retain its enhancement under the same noise~\cite{He:2025ovo}.
We take these prior optimal-measurement results as motivation for adopting the symmetric Dicke state, noting that they do not by themselves establish the noise resilience of the specific parity-readout circuit used here.

Consider $n_q$ qubits coupled to the same local cavity field mode, with collective spin operators
\begin{align}
    J_{\pm} = \sum_{i=1}^{n_q} \sigma_{\pm}^{(i)}, \qquad J_x = \frac{1}{2}(J_+ + J_-).
    \label{eqn:J_pm_def_counting}
\end{align}
The interaction Hamiltonian, Eq.~\eqref{eqn:q_int_H}, for the qubit array can be rewritten as
\begin{align}
    H_{\rm int} = \delta\left( e^{i\alpha}J_+ + e^{-i\alpha}J_-  \right), \qquad 0<t<\tau ,
    \label{eqn:H_int_counting}
\end{align}
where $\delta$ is the approximately constant resonant single-qubit drive amplitude during the interrogation.
Preparing the sensor register in a symmetric Dicke state $|J,m\rangle$ with $J=n_q/2$, whose preparation and collective readout circuit is detailed in Appendix~\ref{app:Ent_Enh}, the interaction operates as
\begin{align}
    U_{\rm sig}^{\otimes n_q}
    \simeq
    I
    -
    i\delta\tau
    \left(
    e^{i \alpha}J_+
    +
    e^{-i \alpha}J_-
    \right).
    \label{eqn:U_sig_counting}
\end{align}
The absorption amplitude for the transitions $|J,m\rangle \rightarrow |J,m+1\rangle$ and $|J,m\rangle \rightarrow |J,m-1\rangle$ is thus enhanced by the matrix elements
\begin{align}
    \langle J, m+1|J_+|J, m\rangle = \sqrt{(J-m)(J+m+1)}, \\
    \langle J, m-1|J_-|J, m\rangle = \sqrt{(J+m)(J-m+1)}.
    \label{eqn:J_matrix_elements_counting}
\end{align}
The signal probability is therefore enhanced by
\begin{align}
    n_{D}(J,m) & = (J-m)(J+m+1) + (J+m)(J-m+1) \nonumber \\
    & = 2(J-m)(J+m) + 2J
    \label{eqn:nD_Jm_counting}
\end{align}
relative to the single-qubit transition probability.
For the symmetric Dicke state with $m=0$, this gives
\begin{align}
    n_D = \frac{1}{2}n_q^2 + n_q .
    \label{eqn:nD_m0_counting}
\end{align}
The linearized collective evolution requires the total transition probability to remain small,
\begin{align}
    p_{\rm sig}^{({\rm Dicke})} = n_D\,\delta^2\tau^2 \ll 1.
\end{align}
The corresponding ideal collective signal count is
\begin{align}\label{eqn:Dicke_Sig}
    N_{\rm sig}^{\rm Dicke} = \left( \frac{n_q^2}{2}+ n_q \right) \, N_{\rm shot} \, p_{\rm sig}.
\end{align}

For the special case $m=0$, the interaction generates the orthogonal states $|J,1\rangle$ and $|J,-1\rangle$ with amplitudes proportional to $e^{i\alpha}$ and $e^{-i\alpha}$, respectively.
Because these states are orthogonal, their probabilities add incoherently and the phase dependence cancels.
Consequently, unlike the GHZ protocol, the Dicke response is independent of the drive phase $\alpha$, and no additional phase-averaging factor appears (see Appendix~\ref{app:Ent_Enh}).

This enhancement is an absorption-matrix-element effect, analogous to Dicke superradiance and superabsorption~\cite{Dicke:1954zz, Higgins:2014cxq}: it requires only a well-controlled relative phase across the qubits within each interrogation shot, not a macroscopic phase stable over the full integration. 
For the collective readout, the practical gain is reduced by state-preparation errors, amplitude damping, dephasing, imperfect mode matching, and readout errors. 
The symmetric Dicke state tolerates amplitude damping better than a GHZ register under excitation-number-resolving readout~\cite{He:2025ovo}, but it does not eliminate these error sources.

\begin{figure}[htbp] 
    \centering
    \includegraphics[width=0.95\columnwidth]{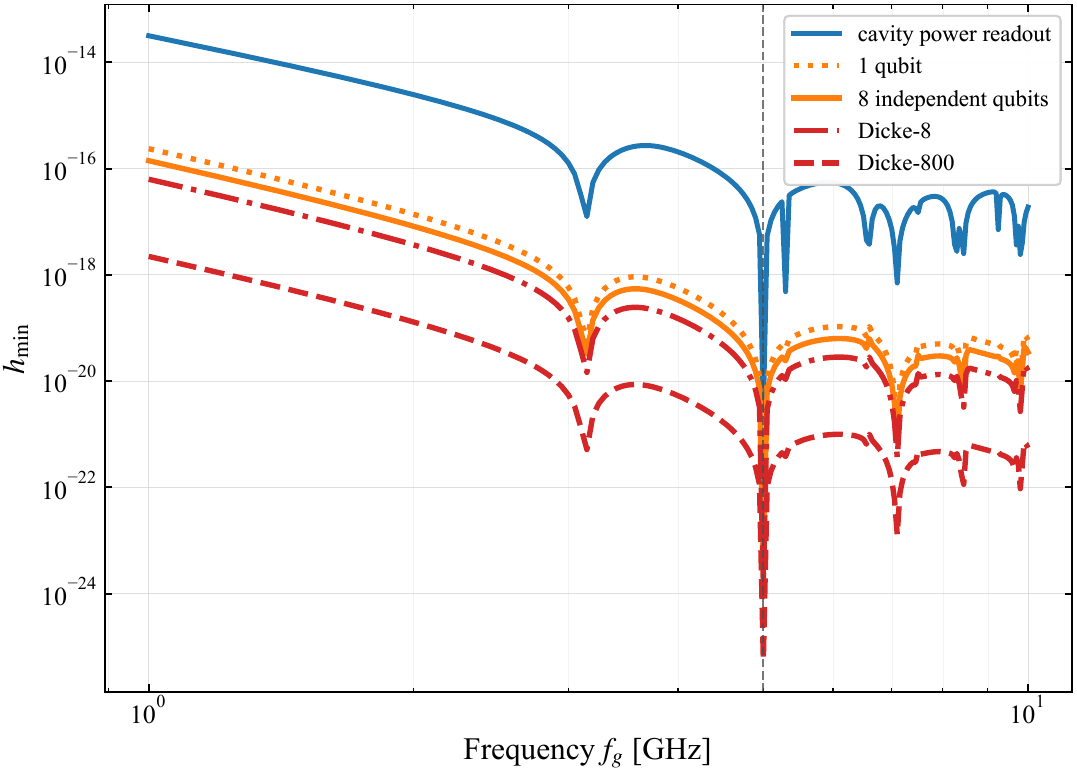} 
    \caption{
    Same hardware, prescribed-field (no-backaction) benchmark, and $5\sigma$ threshold as Figure~\ref{fig:Sensitivity_LR}.
    But for a single fixed $L=R$ cavity of size ($L = R = 6.67 {\rm cm}$) so that the $\mathrm{TE}_{212}$ mode resonates at $5\,\mathrm{GHz}$ (vertical dashed line), with the GW and qubit frequencies scanned over the displayed range. 
    Holding the geometry fixed produces resonant dips at each $\mathrm{TE}_{2nl}$ mode (refer to Figure~\ref{fig:detuned_E_Field}), deepest at $\mathrm{TE}_{212}$.
    }
    \label{fig:Sensitivity_5GHz} 
\end{figure}

\subsection{Qubit readout sensitivity}
\label{sec:Qubit_Sigma}

The statistical significance is estimated as
\begin{align}
    \sigma = \frac{N_{\rm sig}}{\sqrt{N_{\rm bkg}}},
    \label{eqn:sigma_counting}
\end{align}
where $N_{\rm sig}$ is given by Eq.~\eqref{eqn:q-ind_sig}, or Eq.~\eqref{eqn:Dicke_Sig}.
For independent readout we take $N_{\rm bkg} = N_{\rm shot}\,n_q\,p_{\rm bkg}$.
For the ideal collective-state benchmarks, including the Dicke protocols, the background depends on the specific state-preparation and collective-readout circuit. 
In the sensitivity plots below, we therefore use an effective illustrative dark-count bookkeeping model,
\begin{align}
    N_{\rm bkg}
    =
    N_{\rm shot}\,n_q\,p_{\rm bkg},
\end{align}
for both independent and collective protocols. 
This prescription is used only to define a common benchmark comparison and is not intended as a device-level noise model.

The corresponding effective per-shot background probability is modeled as
\begin{align}
    p_{\rm bkg} =
    \begin{cases}
        1-(1-p_{\rm th})(1-p_{\rm ro}),
        & \text{indep.}, \\
        1-(1-p_{\rm th})(1-p_{\rm op})
        \left(1-\dfrac{p_{\rm ro}}{n_q}\right),
        & \text{collective}.
    \end{cases}
    \label{eqn:p_bkg_cases}
\end{align}
Here $p_{\rm th}=1/[\exp(\omega_g/T_{\rm sys})+1]$ is the effective single-qubit thermal excitation probability, $p_{\rm ro}=0.1\%$ is the readout-error parameter, and $p_{\rm op}=0.1\%$ is the effective state-preparation/operation-error parameter used for the benchmark. 
The factor $p_{\rm ro}/n_q$ in the collective case is a bookkeeping convention for the single-ancilla parity readout and should not be interpreted as an independently calibrated per-qubit error rate.

\begin{figure*}[!ht]
    \centering
    \includegraphics[width=0.95\textwidth]{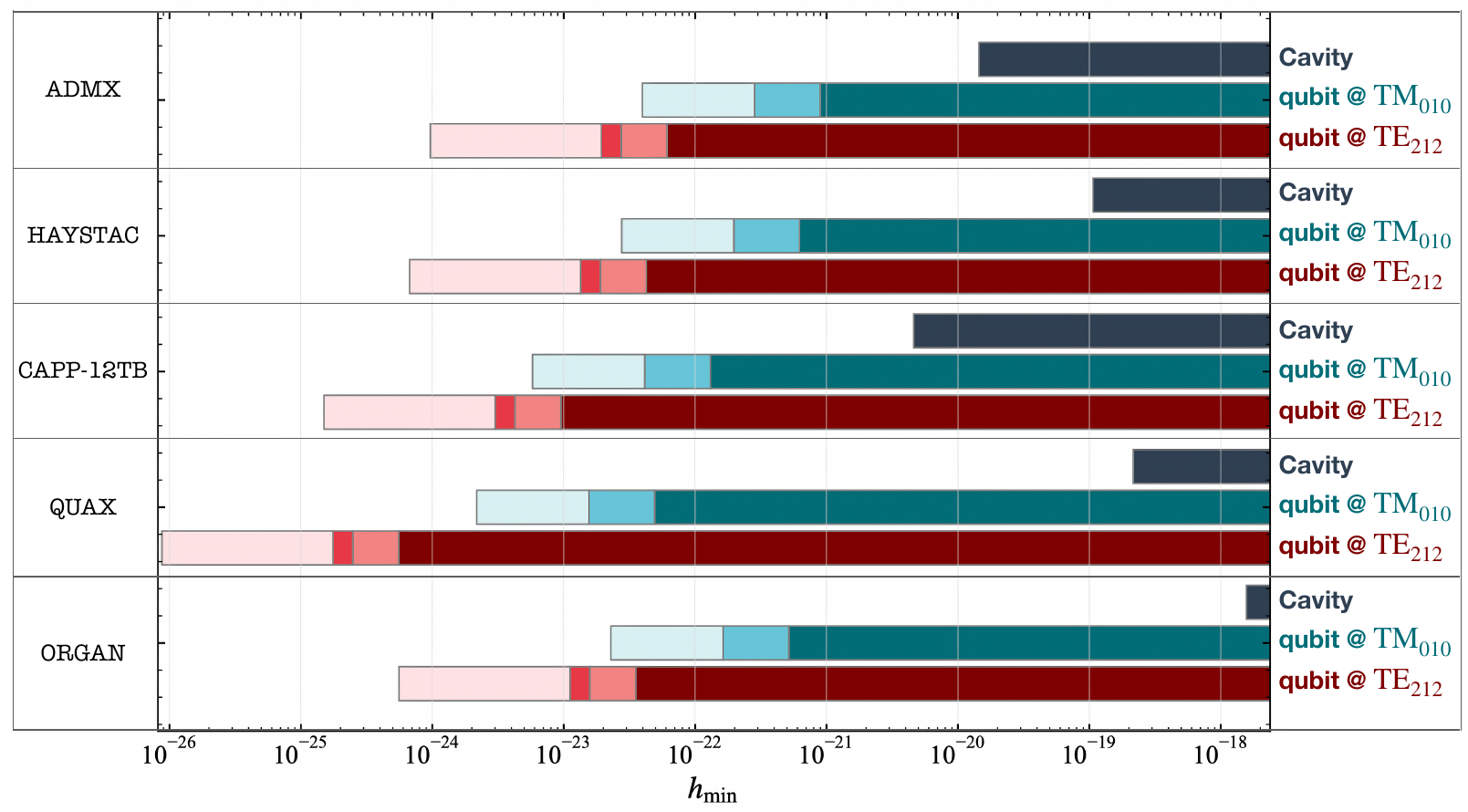}
    \caption{
        $5\sigma$-benchmark projected strain reach $h_{\rm min}$ for representative parameters from five operating haloscope experiments: ADMX~\cite{ADMX:2020hay}, HAYSTAC~\cite{HAYSTAC:2018rwy}, CAPP-12TB~\cite{CAPP:2024dtx}, QUAX~\cite{QUAX:2024fut}, and ORGAN~\cite{Quiskamp:2022xur}. 
        The orientation averages are evaluated for a plus-polarized GW, where $(\theta,\phi)$ specify the cavity-axis orientation and $\psi$ is the rotation of the cylindrical cavity about its own axis; an explicit $h_\times$ calculation is redundant because of the cylindrical azimuthal symmetry.
        The cavity-power benchmark uses the geometry-specific modal RMS overlap $\eta_{\rm cav,rms}(R,L)$ of Eq.~\eqref{eqn:eta_sky_rms}, while each qubit benchmark uses the geometry-specific local response ${\cal R}_{q,\rm rms}(R,L)$ of Eq.~\eqref{eqn:Rq_sky_rms}, both listed in Table~\ref{tab:haloscope_parameters}.
        The all-sky qubit bars assume co-aligned-calibrated local frames; general incidence requires a site-dependent phase recalibration.
        All curves use the same $5\sigma$ detection criterion as Sec.~\ref{sec:Signal_Detection}. 
        Dark navy bars denote cavity-power readout, evaluated for the ${\rm TM}_{010}$ mode.
        For ${\rm TM}_{010}$, the light-cyan, sky-blue, and teal bars show the 100-qubit Dicke, 100 independent-qubit, and single-qubit protocols. 
        For ${\rm TE}_{212}$, the very-light-pink, salmon, bright-red, and dark-red bars show the 800-qubit Dicke, 800 independent-qubit, 8-qubit Dicke, and 8 independent-qubit protocols. The 800-qubit results use the global-register benchmark of Appendix~\ref{app:Ent_Enh}. 
        The experimental parameters are detailed in Appendix~\ref{app:hal_param}.
        The qubit results are prescribed-field (no-backaction) benchmarks that neglect cavity--qubit hybridization and device noise, and therefore should not be interpreted as projected sensitivities of the experiments in their current configurations.
    }
    \label{fig:haloscope_application}
\end{figure*}

This illustrative model does not include register-size-dependent accumulation of gate errors, collective thermal-photon absorption, amplitude damping during the interrogation, or correlated readout errors.
The related background probability is an approximate estimate based on the reported gate fidelities of current superconducting quantum processors. 
Typical IBM superconducting qubits achieve single-qubit gate fidelities exceeding 99.9\% and two-qubit gate fidelities above 99\%, depending on the processor generation and calibration~\cite{Kim:2023bwr, IBMQuantumHardware}. 
A realistic estimate of the background probability requires a full device-specific noise model, including gate errors, decoherence, crosstalk, and readout errors, and should therefore be obtained through dedicated noise simulations or direct experimental characterization.

The minimum detectable strain $h_{\min}$ is obtained by imposing the threshold condition $\sigma=\sigma_*=5$ and solving for the strain through $p_{\rm sig}\propto\delta^2\tau_{\rm eff}^2\propto h^2$.
This yields the protocol-dependent scalings
\begin{align}
    h_{\min}^{\rm q}
    &
    \propto
    \frac{
    \sigma_*^{1/2} p_{\rm bkg}^{1/4}
    }{
    {\cal R}_q  B_0  Q 
    }
    \frac{1}{n_q^{1/4}}
    \frac{1}{C^{1/2}}
    \frac{1}{d}
    \frac{1}{\omega_g^{3/2}}
    \frac{1}{V_{\rm cav}^{1/3}}
    \frac{1}{\tau_{\rm eff}^{3/4} t_{\rm int}^{1/4}} \nonumber \\
    &
    \propto
    \frac{
    \sigma_*^{1/2} p_{\rm bkg}^{1/4}
    }{
    {\cal R}_q  B_0  Q 
    }
    \frac{1}{n_q^{1/4}}
    \frac{1}{C^{1/2}}
    \frac{1}{d}
    \frac{1}{\omega_g^{1/2}}
    \frac{1}{\tau_{\rm eff}^{3/4} t_{\rm int}^{1/4}}
    \label{eqn:hmin_q_counting}
\end{align}
for independent-qubit readout, and
\begin{align}
    h_{\min}^{\rm Dicke}
    & 
    \propto
    \frac{
    \sigma_*^{1/2} p_{\rm bkg}^{1/4}
    }{
    {\cal R}_q B_0  Q  
    }
    \frac{n_q^{1/4}}{(n_q^2+2n_q)^{1/2}}
    \frac{1}{C^{1/2}}
    \frac{1}{d}
    \frac{1}{\omega_g^{3/2}}
    \frac{1}{V_{\rm cav}^{1/3}}
    \frac{1}{\tau_{\rm eff}^{3/4} t_{\rm int}^{1/4}} \nonumber \\
    & 
    \propto
    \frac{
    \sigma_*^{1/2} p_{\rm bkg}^{1/4}
    }{
    {\cal R}_q B_0  Q  
    }
    \frac{n_q^{1/4}}{(n_q^2+2n_q)^{1/2}}
    \frac{1}{C^{1/2}}
    \frac{1}{d}
    \frac{1}{\omega_g^{1/2}}
    \frac{1}{\tau_{\rm eff}^{3/4} t_{\rm int}^{1/4}}
    \label{eqn:hmin_Dicke_counting}
\end{align}
computed for a single global register where the cavity volume $V_{\rm cav}$ is scaled to $\omega^{-3}$ for the second line.
For the monochromatic benchmark $\tau_{\rm eff}\simeq\tau_q$ is frequency independent, so for the $L=R$ cavity of Figure~\ref{fig:Sensitivity_LR} (where the geometric factor  ${\cal R}_q^{\rm peak} \simeq 0.249$ for ${\rm TE}_{212}$ is fixed as the cavity is rescaled) the plotted reach reduces to the closed form $h_{\min}^{\rm q}(f)\propto\omega_g^{-1/2}\,p_{\rm bkg}(f)^{1/4}$, with $p_{\rm bkg}(f)$ carrying the thermal frequency dependence through $p_{\rm th}(f)=1/(e^{\omega_g/T_{\rm sys}}+1)$ at $T_{\rm sys}=0.1\,\mathrm{K}$. 
The reach therefore improves monotonically with frequency, and its falloff is steeper than the explicit $\omega_g^{-1/2}$ factor alone because the thermal background $p_{\rm bkg}(f)$ also decreases as the mode frequency rises above the thermal scale.
Figures~\ref{fig:Sensitivity_LR} and~\ref{fig:Sensitivity_5GHz} evaluate these threshold-strain scalings against frequency for the five protocols. 
Figure~\ref{fig:Sensitivity_LR} rescales the cavity at each point so that the $\mathrm{TE}_{212}$ resonance tracks $f_g$, whereas Figure~\ref{fig:Sensitivity_5GHz} holds a single $5\,\mathrm{GHz}$ cavity fixed and scans the GW and qubit frequencies, which is what produces the resonant dips at the neighboring $\mathrm{TE}_{2nl}$ modes. 
In both cases the local qubit readout improves on the excess-power baseline, the collective protocols improve on independent qubits, and the Dicke benchmark is the most sensitive of the five.
The collective $\sqrt{n_D/n_q}$ gain and the $|m|=2$ mode geometry are the contributions specific to this proposal.

\section{Comparison with Axion DM}
\label{sec:axion_DM}

The cavity-based GW search is closely related to axion haloscopes, which use the same combination of a strong static field and a high-$Q$ resonant cavity~\cite{Berlin:2021txa, Kim:2025izt}. 
The two searches differ in which cavity mode responds, and this difference can be traced to the structure of the interaction at the Lagrangian level.
 
For the axion, the relevant part of the interaction Lagrangian is
\begin{align}\label{eqn:axion_lagrangian}
    \mathcal{L}_{\rm int}
    =
    -\frac{1}{8}g_{a\gamma\gamma}a\,
    \epsilon_{\mu\nu\rho\sigma}
    F^{\mu\nu}F^{\rho\sigma}
    =
    g_{a\gamma\gamma}a\,\mathbf{E}\cdot\mathbf{B},
\end{align}
where $g_{a\gamma\gamma}=(\alpha/2\pi)(C_{a\gamma\gamma}/f_a)$, $f_a$ is the axion decay constant and $C_{a \gamma \gamma}$ is axion--photon anomaly coupling, which depends on the model. 
In a static external field $\mathbf{B}=B_0\hat{\mathbf z}$, varying with respect to $\mathbf{A}$ gives an effective source in Amp\`ere's law,
\begin{align}\label{eqn:axion_current}
    \mathbf{J}_{\rm eff}\simeq g_{a\gamma\gamma}\,\dot a\,\mathbf{B}_0 \, .
\end{align}
This source is approximately uniform, axial, and azimuthally symmetric. 
It therefore overlaps with modes having $E_z\neq0$ and $m=0$, most notably the $\mathrm{TM}_{010}$ mode used in conventional axion haloscopes. 
By contrast, TE modes have $E_z=0$, while modes with $m\neq0$ vanish under the azimuthal overlap integral.
The GW interaction [Eq.~\eqref{eqn:ME_interaction}] produces the effective current in Eq.~\eqref{eqn:eff_curr}, whose quadrupolar angular dependence follows from the spin-2 structure in Eq.~\eqref{eqn:selection_rule}. 
For the cylindrical geometry considered here, this structure gives a strong overlap with the $\mathrm{TE}_{212}$ mode. 
Thus, $\mathrm{TM}_{010}$ is the natural mode for a spatially homogeneous scalar axion source, whereas $\mathrm{TE}_{212}$ provides the representative spin-2 GW channel.

Figure~\ref{fig:haloscope_application} evaluates the qubit-based GW readout of Sec.~\ref{sec:Signal_Detection} using representative cavity parameters from five operating haloscope experiments (Table~\ref{tab:haloscope_parameters} in Appendix~\ref{app:hal_param}). 
For each experiment, the published physical geometries are used to calculate the cavity's own ${\rm TM}_{010}$ eigenfrequency according to Eq.~\eqref{eqn:res_freq_exact}. 
The GW reach is then evaluated for the ${\rm TM}_{010}$ and ${\rm TE}_{212}$ modes using the reported experimental parameters, rather than over the published axion-search frequency range. 

For applications in which the GW incident direction is unknown, we characterize the angular response by orientation averages of the response entering each readout channel.
The analysis is restricted to the plus polarization, $h_+$.
The angles $(\theta,\phi)$ specify the orientation of the cavity axis relative to the GW propagation direction, while $\psi$ denotes a rotation of the cavity about its own axis.
Owing to the azimuthal symmetry of the cylindrical cavity, an explicit calculation for $h_\times$ is redundant: its response is related to the $h_+$ response by an azimuthal rotation already contained in the $(\phi,\psi)$ average.

For the cavity-power channel, the relevant quantity is the RMS modal overlap
\begin{align}
    \eta_{\rm cav,rms}
    \equiv
    \left[
    \frac{1}{8\pi^2}
    \int d\Omega\int_0^{2\pi} d\psi\,
    \left|\eta_{\rm deg}(\theta,\phi,\psi)\right|^2
    \right]^{1/2}.
    \label{eqn:eta_sky_rms}
\end{align}
Distinct cavity eigenmodes are orthogonal under the cavity-volume inner product, so the stored energy and the cavity power contain no cross terms between modes. 
For the local-qubit channel, the relevant quantity is not the modal norm but the local electric-field response already defined in Eq.~\eqref{eq:Ebar_qubit},
\begin{align}
    {\cal R}_{q,\rm rms}
    \equiv
    \left[
    \frac{1}{8\pi^2}
    \int d\Omega\int_0^{2\pi} d\psi\,
    \left|{\cal R}_q(\theta,\phi,\psi)\right|^2
    \right]^{1/2}.
    \label{eqn:Rq_sky_rms}
\end{align}
The $m=\pm2$ cavity modes are orthogonal under the same inner product, but the qubit probes their coherent local-field sum at $\mathbf{x}_q$, so their interference is retained inside ${\cal R}_q$ before the angular average. 
We evaluate both averages with the actual sizes $(R,L)$ of each benchmark cavity, holding the register at fixed normalized body coordinates $(x/R,y/R,z/L)$ and its capacitive axis fixed in the local cavity frame. 
For the qubit benchmarks we take $\tau_q=100~\mu\mathrm{s}$. 
The cavity-power channel uses $\eta_{\rm cav,rms}(R,L)$ of Eq.~\eqref{eqn:eta_sky_rms} and the local-qubit channel uses ${\cal R}_{q,\rm rms}(R,L)$ of Eq.~\eqref{eqn:Rq_sky_rms}. 
The experimental setup paramters are listed in Table~\ref{tab:haloscope_parameters}. 
We introduce no additional normalized overlap or peak-response ratio. 
The all-sky averages assume this phase is fixed by a common calibration of the local frames.
For a source at a general sky position the same phase depends on the incidence direction and the detector orientation, so realizing this calibration in practice would require recalibrating the phase for each direction.

The dark bars show the conventional thermal-noise-limited cavity-power sensitivity obtained following Ref.~\cite{Berlin:2021txa}, evaluated for the ${\rm TM}_{010}$ mode.
The light-blue bars show the ${\rm TM}_{010}$-mode qubit sensitivities for a single qubit, 100 independent qubits, and a 100-qubit Dicke-state protocol. 
The red bars show the corresponding ${\rm TE}_{212}$-mode sensitivities for 8 and 800 independent qubits and for 8- and 800-qubit Dicke-state protocols. The 800-qubit benchmarks are evaluated using the global-register prescription described in Appendix~\ref{app:Ent_Enh}.

\section{Conclusion and Discussion}
\label{sec:Conclusion}

We have developed a quantum-sensing paradigm for detecting high-frequency gravitational waves (HFGWs) in the GHz band that exploits the characteristic spin-2 quadrupolar feature of the GW-induced electromagnetic field~\cite{Chen:2022quj, Chen:2024aya}.
Instead of measuring the total cavity power~\cite{Berlin:2021txa, Kim:2025izt}, we place superconducting transmon qubits at the electric-field hot spots of a resonant cylindrical cavity and read the induced field locally.
This approach allows a localized quantum-enhanced extraction of the gravitational signal.

Configuring these local sensors as an entangled register causes the signal probability grow as $n_q^{2}$, so the strain reach scales as $h_{\min}^{\rm Dicke}\propto n_q^{-3/4}$ (Figures~\ref{fig:Sensitivity_LR} and~\ref{fig:Sensitivity_5GHz}).
At $5~\mathrm{GHz}$, for a GW propagating along the cavity axis, an $800$-qubit global register reaches an idealized $h_{\min}\simeq5.6\times10^{-26}$, nearly five orders of magnitude below the cavity-power benchmark, and splitting it into eight local ones at the $\mathrm{TE}_{212}$ hot spots costs only a factor of about $2.8$.
Since the axion coupling $g_{a\gamma\gamma}=(\alpha/2\pi)C_{a\gamma\gamma}/f_a$ exceeds $1/M_{\rm Pl}$ by $\mathcal{O}(10^{4})$ and the source amplitudes differ as well, Figure~\ref{fig:haloscope_application} shows how far bias fields, resonant enhancement, optimal placement, and collective matrix elements close that gap in the geometries considered here.

Translating these idealized, prescribed-field upper limits into deployable hardware demands a rigorous open-quantum-system approach. 
Future projections must dynamically incorporate cavity loss ($\kappa$), qubit decoherence ($\gamma_1, \gamma_\phi$), and protocol-specific gate errors (Appendix~\ref{app:cavity_qubit_hybridization}). 
Furthermore, realizing these hybrid platforms requires overcoming severe engineering hurdles: preserving transmon coherence within intense magnetic bias fields—via next-generation field-resilient designs~\cite{Krause:2021llk, Gunzler:2025spt, PhysRevApplied.15.054001} or spatial shielding~\cite{Kono:2017ynw}—while strictly controlling the cavity quality-factor degradation induced by qubit integration.

Ultimately, we emphasize that this framework lays the foundation for macroscopic networks of quantum-enhanced cavity detectors. 
The directional sensitivity inherent to our local readout dictates that cross-correlating a distributed detector array could reconstruct the polarization and origin of coherent transient sources, as well as isolate stochastic HFGW backgrounds from uncorrelated instrumental noise.
By systematically addressing the identified open-system and hardware challenges, this architecture stands to fundamentally expand the experimental frontier for high-frequency gravitational waves and open a new window onto the early universe and physics beyond the Standard Model.

\acknowledgments

We thank Hajime Fukuda, SungWoo Youn, and Sungmok Lee for helpful discussions and comments.
KK is supported in part by the US DOE under Award No DE-SC0024673.
MP is supported by the National Research Foundation of Korea (NRF) grant funded by the Korea government (MSIT) (No.RS-2025-00564488). 
HY is supported by the National Research Foundation of Korea (NRF) grant funded by the Korea government (MSIT) (No.RS-2025-00564488) and (RS-2024-00340153). 
SCP is supported by the National Research Foundation of Korea (NRF) grant funded by the Korea government (MSIT) (RS-2024-00340153).

\appendix

\section{PD frame metric for a monochromatic plane wave}
\label{app:PD_frame}

This appendix collects the explicit PD frame metric used in Sec.~\ref{sec:Theoretical_Setup} to evaluate the effective current of Eq.~\eqref{eqn:eff_curr}. Following Refs.~\cite{Marzlin:1994ia, Berlin:2021txa}, the metric in the PD frame is
\begin{subequations}\label{eqn:PD_frame}
\begin{align}
    h_{00}
    &=
    -2
    \sum_{n=0}^{\infty}
    \frac{n+3}{(n+3)!}\,
    R_{0k0l,m_{1}\cdots m_{n}}\,
    x^{k}\,x^{l}\,x^{m_{1}}\cdots x^{m_{n}} ,
    \label{eqn:h00_PD}
    \\
    h_{0j}
    &=
    -2
    \sum_{n=0}^{\infty}
    \frac{n+2}{(n+3)!}\,
    R_{0kjl,m_{1}\cdots m_{n}}\,
    x^{k}\,x^{l}\,x^{m_{1}}\cdots x^{m_{n}} ,
    \label{eqn:h0j_PD}
    \\
    h_{ij}
    &=
    -2
    \sum_{n=0}^{\infty}
    \frac{n+1}{(n+3)!}\,
    R_{ikjl,m_{1}\cdots m_{n}}\,
    x^{k}\,x^{l}\,x^{m_{1}}\cdots x^{m_{n}} ,
    \label{eqn:hij_PD}
\end{align}
\end{subequations}
where $R_{\mu \nu \rho \sigma} \propto e^{i \omega_g (t-z)}$ is the Riemann tensor of a monochromatic plane wave traveling along the $\hat{z}$ direction. The indices $i,j,k,l$ denote spatial indices, while $i$ appearing outside an index position denotes the imaginary unit. The Riemann tensor $R_{\mu\nu\rho\sigma}$ is evaluated at the spatial origin $x^i=0$, chosen to coincide with the detector center, while its time dependence is kept explicit. The finite-wavelength factors in Eq.~\eqref{eqn:PD_frame} reduce to unity in the limit $\omega_g \to 0$, reproducing the standard long-wavelength expressions.

Evaluating the Riemann tensor for the TT-gauge plane wave of Eq.~\eqref{eqn:h_TT}, the metric components are obtained exactly as~\cite{Berlin:2021txa}
\begin{widetext}
\begin{subequations}\label{eq:h_PD_exact}
\begin{align}
    h_{00}
    &=
    -\omega_{g}^{2}\,h_{ab}^{\rm TT}\,x^{a}\,x^{b}
    \left[
        -\frac{i}{\omega_{g}\,z}
        + \frac{1 - e^{-i\omega_{g}z}}{(\omega_{g}\,z)^{2}}
    \right] ,
    \label{eq:h00_exact}
    \\
    h_{ij}
    &=
    \omega_{g}^{2}
    \left[
        \bigl(\delta_{iz}\,h_{ja}^{\rm TT} + \delta_{jz}\,h_{ia}^{\rm TT}\bigr)\,z\,x^{a}
        - h_{ij}^{\rm TT}\,z^{2}
        - \delta_{iz}\,\delta_{jz}\,h_{ab}^{\rm TT}\,x^{a}\,x^{b}
    \right]
    \left[
        -\frac{1 + e^{-i\omega_{g}z}}{(\omega_{g}\,z)^{2}}
        - 2i\,\frac{1 - e^{-i\omega_{g}z}}{(\omega_{g}\,z)^{3}}
    \right] ,
    \label{eq:hij_exact}
    \\
    h_{0i}
    &=
    -\omega_{g}^{2}
    \left(
        h_{ia}^{\rm TT}\,z\,x^{a}
        - \delta_{iz}\,h_{ab}^{\rm TT}\,x^{a}\,x^{b}
    \right)
    \left[
        -\frac{i}{2\omega_{g}\,z}
        - \frac{e^{-i\omega_{g}z}}{(\omega_{g}\,z)^{2}}
        - i\,\frac{1 - e^{-i\omega_{g}z}}{(\omega_{g}\,z)^{3}}
    \right] ,
    \label{eq:h0i_exact}
\end{align}
\end{subequations}
\end{widetext}
where $a, b = x, y$ is the transverse index such that $h_{xx} = - h_{yy} = h_+$ and $h_{xy} = h_{yx} = h_{\times}$.

Several features of Eq.~\eqref{eq:h_PD_exact} are worth noting. 
Writing $x=\omega_gz$, the three bracketed functions multiplying $h_{00}$, $h_{ij}$, and $h_{0i}$ approach $1/2$, $1/6$, and $1/3$, respectively, as $x\to0$. They do not individually approach unity.
Including the curvature-dependent prefactors in Eq.~\eqref{eq:h_PD_exact} recovers the standard leading-order PD/Fermi-normal metric~\cite{Maggiore:2007ulw,Marzlin:1994ia}.
The resummed expressions are exact at first order in strain for the monochromatic plane wave and retain finite-wavelength dependence when $\omega_gL_{\rm det}=\mathcal O(1)$.  
The TT amplitudes specify the physical incoming polarization, while the displayed metric components are those of the PD coordinates.

\section{Electromagnetic responses of different modes in cylindrical cavity}
\label{app:mode_expansion}

This appendix collects the cavity-mode functions, their normalization, and the basis conventions used in Sec.~\ref{sec:Electromagnetic_Signal}, and gives the selection rule that singles out the quadrupolar ($m=2$) modes. Throughout we follow the conventions of Ref.~\cite{Berlin:2021txa} and standard cavity electrodynamics~\cite{collin1990field, jackson1999classical, hill2009electromagnetic}.
 
The electric field can be Helmholtz-decomposed into solenoidal modes, $\nabla\cdot\mathbf E_{sn}=0$, and irrotational modes, $\nabla\times\mathbf E_{in}=0$, both satisfying the conducting tangential-field boundary condition.  
The projection is carried out on the curl--curl equation~\eqref{eqn:curlcurl}.
Using the vector identity
$\nabla\times(\nabla\times\mathbf{E})=\nabla(\nabla\cdot\mathbf{E})-\nabla^2\mathbf{E}$
and $\nabla\cdot\mathbf{E}_{sn}=0$, together with the PEC boundary condition
$\hat n\times\mathbf{E}_c=0$ under which
$\int_V (\nabla\times\nabla\times\mathbf{E})\cdot\mathbf{E}_c^* \,d^3x
=\int_V \mathbf{E}\cdot(\nabla\times\nabla\times\mathbf{E}_c)^* \,d^3x$,
the solenoidal projection of Eq.~\eqref{eqn:curlcurl} yields
Eq.~\eqref{eqn:Wave_eqn} directly: the charge-gradient source
$\nabla\rho_{\rm eff}$ never enters the solenoidal sector, and no boundary
term is left unevaluated.
The charge-gradient and longitudinal-current terms drive only the
irrotational sector, related by
$\partial_t\rho_{\rm eff}+\nabla\cdot\mathbf j_{\rm eff}=0$, which receives
no resonant $Q$ enhancement.
 
We use detector-centered coordinates $z\in[-L/2,L/2]$ and define $\zeta\equiv z+L/2\in[0,L]$, with $k_z=l\pi/L$.  
TM modes allow $l=0,1,\ldots$, whereas the electric fields written below for TE modes require $l=1,2,\ldots$. The nominal TE $l=0$ expression vanishes identically.
The transverse wavenumber is $\gamma_{mn}=x_{mn}/R$ for TM modes and $\gamma_{mn}=x'_{mn}/R$ for TE modes.  
In the complex basis $e^{im\phi}$, one consistent overall-phase convention is
\begin{subequations}\label{eqn:TM_mode}
\begin{align}
    E_r^{\rm TM} &= -\frac{k_z}{\gamma_{mn}}\,J'_m(\gamma_{mn}r)\,e^{im\phi}\sin(k_z\zeta),\\
    E_\phi^{\rm TM} &= -\frac{i\,m\,k_z}{\gamma_{mn}^{2}\,r}\,J_m(\gamma_{mn}r)\,e^{im\phi}\sin(k_z\zeta),\\
    E_z^{\rm TM} &= J_m(\gamma_{mn}r)\,e^{im\phi}\cos(k_z\zeta),
\end{align}
\end{subequations}
and the $\mathrm{TE}_{mnl}$ mode functions are
\begin{subequations}\label{eqn:TE_mode}
\begin{align}
    E_r^{\rm TE} &= \frac{\omega_{mnl}\,m}{\gamma_{mn}^{2}\,r}\,J_m(\gamma_{mn}r)\,e^{im\phi}\sin(k_z\zeta),\\
    E_\phi^{\rm TE} &= \frac{i\,\omega_{mnl}}{\gamma_{mn}}\,J'_m(\gamma_{mn}r)\,e^{im\phi}\sin(k_z\zeta),\\
    E_z^{\rm TE} &= 0 ,
\end{align}
\end{subequations}
where $\omega_{mnl}^{2}-k_z^{2}=\gamma_{mn}^{2}$.  
Direct substitution gives $\nabla\cdot\mathbf E^{\rm TM}=\nabla\cdot\mathbf E^{\rm TE}=0$, and the shift $\zeta=z+L/2$ makes the tangential electric field vanish at both endcaps.  
For $m\neq0$ the two signs $\pm m$ form a degenerate orthogonal pair, and the real basis follows from their linear combinations.
Overall constants are fixed by Eq.~\eqref{eqn:mode_condition_2}.
 
Using $\nabla^{2}\mathbf{E}_{mnl}=-\omega_{mnl}^{2}\mathbf{E}_{mnl}$ [Eq.~\eqref{eqn:mode_condition_1}] together with $\nabla^{2}=-(\gamma_{mn}^{2}+k_z^{2})$ on the separable mode functions, the resonant frequencies are
\begin{align}\label{eqn:res_freq_exact}
    \omega_{mnl}=\sqrt{\left(\frac{x_{mn}}{R}\right)^{2}+\left(\frac{l\pi}{L}\right)^{2}}\ \ (\mathrm{TM}),\\
    \omega_{mnl}=\sqrt{\left(\frac{x'_{mn}}{R}\right)^{2}+\left(\frac{l\pi}{L}\right)^{2}}\ \ (\mathrm{TE}),
\end{align}
which is the exact version of the estimate in Eq.~\eqref{eqn:mode_freq_approx}. Representative zeros relevant here are $x_{01}\simeq2.405$, $x'_{11}\simeq1.841$, and $x'_{21}\simeq3.054$, and the $\mathrm{TE}_{212}$ mode corresponds to $m=2$, $n=1$, $l=2$.
 
The overlap factor of Eq.~\eqref{eqn:overlap_factor} is obtained by projecting the GW-induced current onto each mode. When the GW propagation direction, the cavity axis, and the static field $\mathbf{B}_0$ are co-aligned, the effective current carries a definite azimuthal structure $\propto e^{\pm 2i\phi}$, reflecting the spin-2 (quadrupolar) character of the metric perturbation. The azimuthal integral in Eq.~\eqref{eqn:overlap_factor} then reduces to
\begin{align}\label{eqn:selection_rule}
    \int_{0}^{2\pi} d\phi\ e^{\pm 2i\phi}\,e^{-im\phi}\ \propto\ \delta_{m,\,\pm 2} ,
\end{align}
so that only modes with azimuthal index $|m|=2$ have nonvanishing overlap at leading order in this configuration. This is why the main text takes quadrupolar modes to dominate the response, with $\mathrm{TE}_{212}$ as a representative target. For more general orientations the cylindrical symmetry is broken and additional azimuthal components contribute, as discussed in Appendix~\ref{app:rot_response}. 
Numerically, the dimensionless overlap takes values $\eta\sim\mathcal{O}(0.1)$ for the mode choices and geometries considered here.

\begin{figure}[t]
    \centering
     \includegraphics[width=0.99\columnwidth]{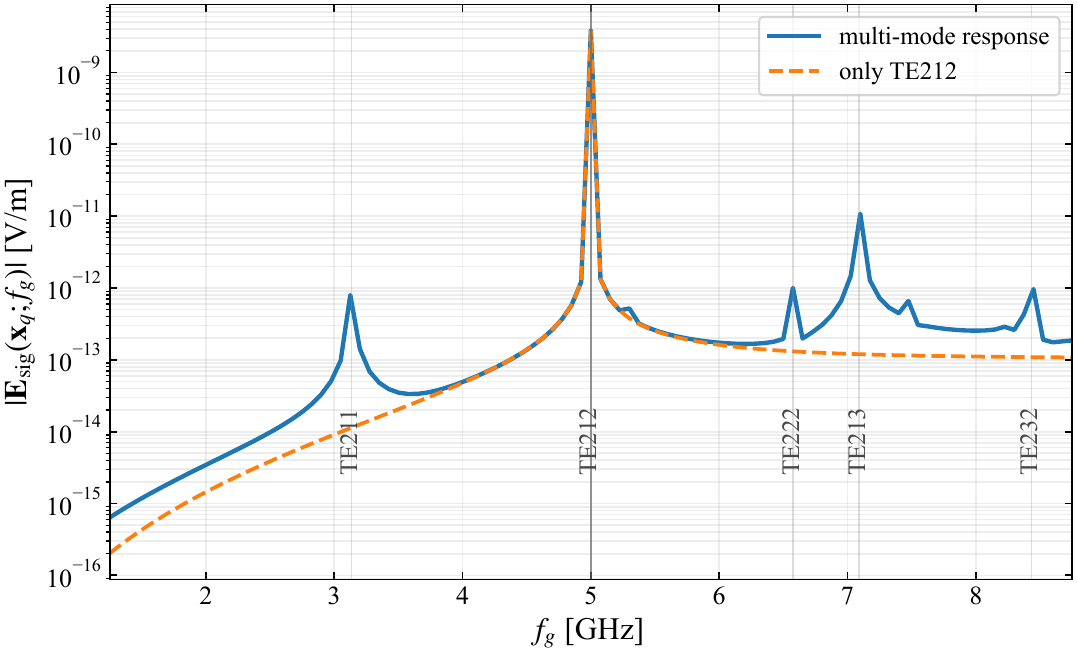}
    \caption{
    Detuned response at the fixed $\mathrm{TE}_{212}$ hot spot for the coaxial benchmark ($\theta=\phi=0$, $h_0 = |h_+|=10^{-23}$, $R=L=6.67~\mathrm{cm}$ so that $f_{\mathrm{TE}_{212}}=5~\mathrm{GHz}$, $Q_c = Q =10^5$).
    The plot compares the coherent $\mathrm{TE}_{2nl}$ multimode sum (solid) with the single $\mathrm{TE}_{212}$ mode (dashed), as the drive frequency $f_g$ is scanned.
    The axis is $|\mathbf{E}_{\rm sig}(\mathbf{x}_q;f_g)|$ at the qubit position.
    Off resonance, the multimode sum picks up the neighboring $\mathrm{TE}_{2nl}$ peaks (labeled).
    }
    \label{fig:detuned_E_Field}
\end{figure}

\begin{figure*}[t]
    \centering
    \begin{subfigure}{0.49\textwidth}
        \centering
        \includegraphics[width=\linewidth]{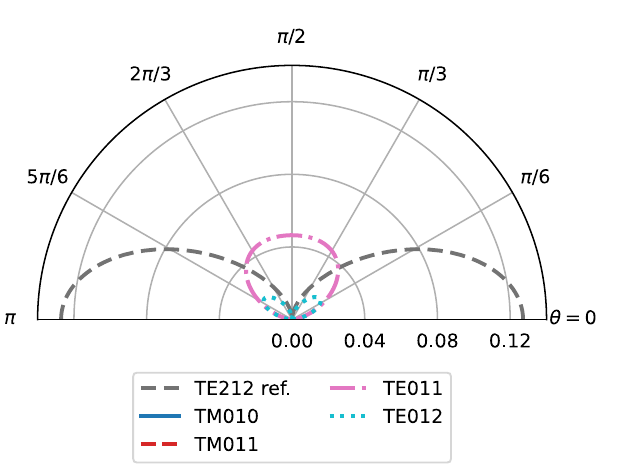}
        \caption{$m=0$ examples}
        \label{fig:m0_mode_theta}
    \end{subfigure}
    \begin{subfigure}{0.49\textwidth}
        \centering
        \includegraphics[width=\linewidth]{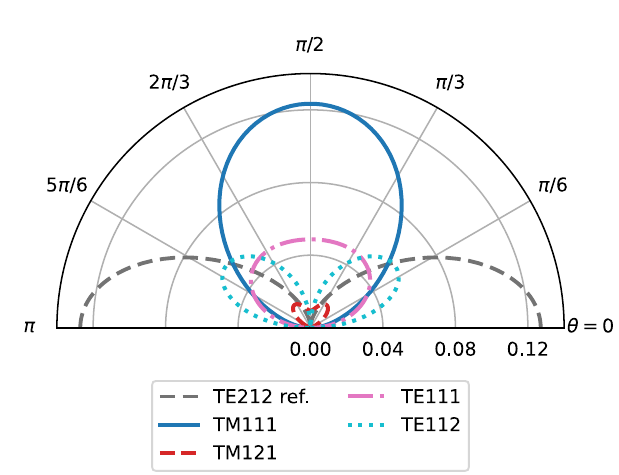}
        \caption{$m=1$ examples}
        \label{fig:m1_mode_theta}
    \end{subfigure}
    \caption{
    Polar-angle dependence $|\eta_{\rm deg}(\theta)|$ of the degenerate-pair overlap [Eq.~\eqref{eqn:eta_deg}] at fixed $\phi=0$, for representative $m=0$ modes (a) and $m=1$ modes (b), with the $\mathrm{TE}_{212}$ reference (gray dashed) in both panels. The non-$\mathrm{TE}_{212}$ modes gain appreciable overlap only once the cavity is tilted off-axis.
    The $m=0$ TM overlaps vanish identically in the $\phi=0$ plane by symmetry, so the ${\rm TM}_{010}$ and ${\rm TM}_{011}$ curves draw no visible line in panel (a); see Figure~\ref{fig:diff_mode_phi}.
    }
    \label{fig:diff_mode_theta}
\end{figure*}

\begin{figure*}[t]
    \centering
     \includegraphics[width=0.95\textwidth]{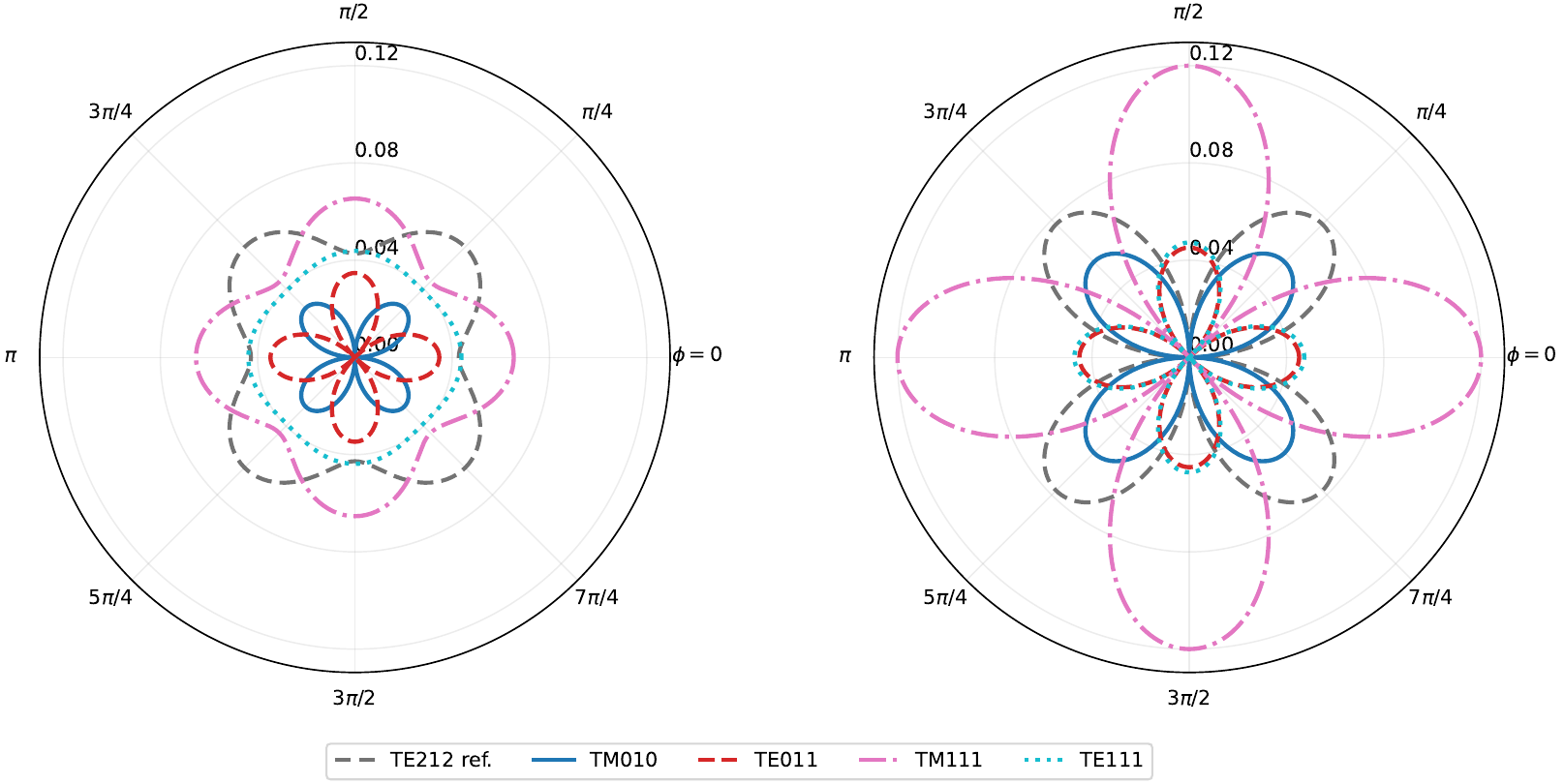}
    \caption{
    Azimuthal dependence $|\eta_{\rm deg}(\phi)|$ at fixed tilt $\theta=\pi/4$ (left) and $\theta=\pi/2$ (right), for $\mathrm{TE}_{212}$ (reference), $\mathrm{TM}_{010}$, $\mathrm{TE}_{011}$, $\mathrm{TM}_{111}$, and $\mathrm{TE}_{111}$, with $|\eta_{\rm deg}|$ as in Figure~\ref{fig:diff_mode_theta}. The number of lobes follows the mode's azimuthal index.
    }
    \label{fig:diff_mode_phi}
\end{figure*}

\subsection{Detuned frequency response}
\label{app:detune response}

In the main text the signal field Eq.~\eqref{eqn:E_sig} is quoted on resonance, $\omega_g\simeq\omega_c$. Here we record the response for a general detuning, which defines the detector transfer function. Returning to the mode equation~\eqref{eqn:Wave_eqn} with a monochromatic source $\mathbf{j}_{\rm eff}=e^{i\omega_g t}\mathbf{J}_{\rm eff}$, the steady-state field is
\begin{align}\label{eqn:detuned_field}
    \mathbf{E}_{\rm sig}(\mathbf{x},t)
    & =\sum_{c\in\mathcal{M}} T_c(\omega_g)\,
    \frac{\int_{V_{\rm Cav}} d^3\mathbf{x}\,\mathbf{E}^{*}_{c}\cdot\mathbf{J}_{\rm eff}}{\int_{V_{\rm Cav}} d^3\mathbf{x}\,|\mathbf{E}_{c}|^{2}}\,
    \mathbf{E}_c(\mathbf{x})\,e^{i\omega_g t},\\
     \quad \text{where} & \quad T_c(\omega_g) \equiv\frac{-i\,\omega_g}{\omega_c^{2}-\omega_g^{2}+i\,\omega_g\omega_c/Q_c} .
\end{align}
Here $\mathcal{M}$ denotes a general set of modes with mode-dependent
resonant frequencies $\omega_c$ and quality factors $Q_c$; the exactly
degenerate $\pm m$ pair of the main text is the special case
$\mathcal{D}\subset\mathcal{M}$ with common $\omega_c$.
On resonance, $\omega_g=\omega_c$, the denominator becomes $i\,\omega_g\omega_c/Q_c$ and $T_c\to-Q_c/\omega_g$, reproducing Eq.~\eqref{eqn:E_sig}. Near a given resonance the transfer function reduces to a Lorentzian,
\begin{align}\label{eqn:Tc_lorentzian}
    |T_c(\omega_g)|^{2}\simeq\frac{1}{4(\omega_g-\omega_c)^{2}+(\omega_c/Q_c)^{2}} ,
\end{align}
with full width at half maximum $\Delta\omega=\omega_c/Q_c$, i.e. $\Delta f=f_c/Q_c$, consistent with the cavity-linewidth contribution to the effective bandwidth in Eq.~\eqref{eq:cavity_bandwidth}.

When several modes lie close in frequency, the response at a fixed observation point is the superposition of the corresponding Lorentzians, so the off-resonant tails of neighboring modes add to the nominal target mode. Figure~\ref{fig:detuned_E_Field} shows this multimode response evaluated at the fixed $\mathrm{TE}_{212}$ hot spot. The coherent $\mathrm{TE}_{2nl}$ sum (solid) coincides with the single-mode $\mathrm{TE}_{212}$ pattern (dashed) on resonance, and develops additional peaks at the neighboring $\mathrm{TE}_{2nl}$ eigenfrequencies off resonance.

\subsection{Rotational dependence}
\label{app:rot_response}

The overlap factor depends on the relative orientation of the incoming GW, the cavity axis, and the static field. We use the two coordinate systems of Figure~\ref{fig:Cavity_Rotation}: a global frame $(x,y,z)$ in which the GW propagates along $\hat z$ with $h_{\mu\nu}\propto e^{i\omega_g(t-z)}$, and a local frame $(x',y',z')$ adapted to the cavity, with the cavity axis and $\mathbf{B}_0$ along $\hat z'$. The two frames are related by a rotation $R(\theta,\phi)\in SO(3)$, where $\theta$ is the angle between $\hat z$ and $\hat z'$ and $\phi$ specifies the orientation of the cavity axis about $\hat z$.
 
To evaluate the overlap factor for an arbitrary orientation, the effective current $\mathbf{J}_{\rm eff}$, which is most simply expressed in the global (GW) frame, is transformed into the local cavity frame through $R(\theta,\phi)$ and projected onto the mode functions before the volume integration. Because the GW-induced current transforms as a spin-2 quantity, a rotation of the cavity axis away from the GW propagation direction mixes the $m=\pm2$ helicity components of the source. This breaks the cylindrical symmetry and, in general, allows the source to overlap with cavity modes of azimuthal index other than the co-aligned $|m|=2$ selection of Eq.~\eqref{eqn:selection_rule}.
 
For a mode with azimuthal index $m\neq0$ the source excites the degenerate $(\pm m)$ pair, and we quote the pair-summed overlap
\begin{align}\label{eqn:eta_deg}
    |\eta_{\rm deg}|^{2}=|\eta_{+}|^{2}+|\eta_{-}|^{2} ,
\end{align}
which reduces to the single-mode result for $m=0$. Figure~\ref{fig:diff_mode_theta} shows the polar-angle dependence $|\eta_{\rm deg}(\theta)|$ at fixed $\phi=0$ for representative $m=0$ and $m=1$ modes, with the $\mathrm{TE}_{212}$ reference repeated in both panels. The $\mathrm{TE}_{212}$ overlap is maximal when the cavity axis is aligned or anti-aligned with the GW propagation direction ($\theta=0,\pi$) and is suppressed near the transverse orientation ($\theta=\pi/2$), while the $m=0$ and $m=1$ modes acquire appreciable overlap only once the cavity is tilted. Figure~\ref{fig:diff_mode_phi} shows the residual azimuthal dependence $|\eta_{\rm deg}(\phi)|$ at fixed $\theta=\pi/4$ and $\theta=\pi/2$, whose multilobed structure reflects the projection of the cavity mode onto the GW polarization basis. An analogous angular analysis was carried out in Ref.~\cite{Berlin:2021txa}.

\section{Qubit system for sensing}
\label{app:Q_Sys}

This appendix records the coupling of the local GW-induced microwave field to a transmon qubit and to a collective qubit register, following the direct-excitation treatment of transmon sensors developed for weak microwave signals~\cite{Chen:2022quj, Chen:2024aya}.

\subsection{Signal interaction Hamiltonian}
\label{app:Int_Ham}

The transmon couples to the gravitationally induced signal field through a dipole-type interaction. 
Restricting to the qubit's two-level subspace, the Hamiltonian takes the form~\cite{Chen:2022quj, Chen:2023swh, Chen:2024aya}
\begin{align}\label{eqn:qubit_Ham}
    H = \omega_q |e\rangle \langle e| + 2 \delta \cos(\omega_g t - \alpha) (|e\rangle \langle g| + |g\rangle \langle e|),
\end{align}
where $\omega_q$ is the qubit transition frequency, $\omega_g$ is the gravitational-wave frequency, $\delta$ is the coupling strength set by the local mode amplitude at the qubit's position, and $\alpha$ is a position-dependent phase inherited from the cavity mode function. 
The treatment below assumes the qubit evolves unitarily under $H$, which holds over a coherence time $\tau \simeq \min(\tau_q, \tau_{\rm GW})$ set by whichever decoheres first, the qubit itself or the phase coherence of the signal~\cite{Chen:2024aya}.
For the monochromatic benchmark, $\tau_{\rm GW}\gg\tau_q$ and the cavity is assumed to be continuously driven into its steady state before the interrogation.

To remove the fast carrier at $\omega_q$, we write the state as
\begin{align}
    |\psi(t)\rangle = \psi_g(t)|g\rangle + \psi_e(t) e^{-i \omega_q t}|e\rangle,
\end{align}
absorbing the free qubit evolution into the excited-state amplitude~\cite{Chen:2024aya}. 
Substituting above expression into the Schr\"odinger equation
\begin{align}
    i\frac{d}{dt} |\psi(t)\rangle = H | \psi(t) \rangle
\end{align}
and collecting the coefficients of the basis states $|g\rangle$ and $|e\rangle$ separately, we obtain
\begin{align}
    i
    \begin{pmatrix}
        \dot{\psi}_g \\
        \dot{\psi}_e
    \end{pmatrix}
    = 
    2 \delta \cos(\omega_g t - \alpha)
    \begin{pmatrix}
        e^{-i \omega_q t}\psi_e \\
        e^{+i \omega_q t}\psi_g
    \end{pmatrix}.
\end{align}
Expanding the cosine and dropping the terms oscillating at $\omega_q + \omega_g$ (the rotating-wave approximation) leaves only the near-resonant terms,
\begin{align}
    i \dot{\psi}_ g & = \delta e^{-i \Delta_{qg} t - i \alpha} \psi_e, \\
    i \dot{\psi}_ e & = \delta e^{i \Delta_{qg} t + i \alpha} \psi_g,
\end{align}
with detuning $\Delta_{qg} = \omega_q - \omega_g \simeq 0$ for a qubit tuned near the signal frequency. 
At exact resonance this reduces to a time-independent two-level problem,
\begin{align}
    i
    \begin{pmatrix}
        \dot{\psi_g} \\
        \dot{\psi_e}
    \end{pmatrix}
    =
    \delta
    \begin{pmatrix}
        0 & e^{- i \alpha} \\
        e^{i \alpha} & 0
    \end{pmatrix}
    \begin{pmatrix}
        \psi_g \\
        \psi_e
    \end{pmatrix},
\end{align}
whose exact solution is a Rabi rotation,
\begin{align}
    \begin{pmatrix}
        \psi_g(t) \\
        \psi_e(t)
    \end{pmatrix}
    =
    \underbrace{
    \left[
    \cos(\delta t) \mathbb{I} - i \sin(\delta t) 
    \begin{pmatrix}
        0 & e^{- i \alpha} \\
        e^{i \alpha} & 0
    \end{pmatrix}
    \right]}_{\equiv U_{\rm sig}^{(+)}(t)}
    \begin{pmatrix}
        \psi_g(0) \\
        \psi_e(0)
    \end{pmatrix}.
\end{align}
Unlike the perturbative expansion of $U_{\rm DM}(t)$ to first order in the coupling used for the weak, always-on axion signal~\cite{Chen:2024aya}, the protocols considered here apply the signal Hamiltonian for a controlled duration $t$, so the full resonant solution is retained without linearizing in $\delta$.

The phase $\alpha$ is not uniform across the cavity. 
Qubits placed at different hot spots of the same mode see it shifted by $\pi$, as set by the mode functions derived in Appendix~\ref{app:mode_expansion}.
Repeating the derivation with $\alpha \to \alpha + \pi$ flips the sign of the off-diagonal coupling and gives
\begin{align}
    U_{\rm sig}^{(-)}(t) = \left[\cos(\delta t) \mathbb{I} + i \sin(\delta t) 
    \begin{pmatrix}
        0 & e^{- i \alpha} \\
        e^{i \alpha} & 0
    \end{pmatrix}
\right] = \left(U_{\rm sig}^{(+)}(t)\right)^\dagger.
\end{align}
$U_{\rm sig}^{(-)}$ is the Hermitian conjugate of $U_{\rm sig}^{(+)}$, so the two evolutions are related by a single-qubit $Z$ gate. 
Conjugating $U_{\rm sig}^{(-)}$ with $Z$ restores the original sign structure.
It is therefore sufficient to track one convention, $U_{\rm sig} \equiv U^{(+)}_{\rm sig}$, and recover the other hot spot response through
\begin{align}
    U_{\rm sig} = Z\,U_{\rm sig}^{(-)}\,Z.
\label{eqn:U_phase}
\end{align}
Given the mode patterns shown in Figure~\ref{fig:TE212_SignalField}, we use Eq.~\eqref{eqn:U_phase} to define a local qubit frame at each electric-field hot spot.
At sites where the mode phase differs by $\pi$, the frame definition includes conjugation by a local $Z$ gate.
The Dicke state and the collective readout below are defined in these local frames.

\subsection{Open-system qubit detection near cavity resonance}
\label{app:cavity_qubit_hybridization}

The qubit-readout analysis of Sec.~\ref{sec:qubit_readout} assumes that the GW-induced cavity field is generated independently of the detector qubit, adopting a prescribed-field (no-backaction) benchmark in which the bare-cavity field is applied as an external drive.
Near simultaneous cavity--qubit resonance, however, a cavity photon and a qubit excitation form polariton states, and their dynamics must be treated together with dissipation.
We first identify the hybrid states generated by the coherent Hamiltonian and then formulate the open-system evolution required for a realistic qubit-detection probability.

Restricting the dynamics to the cavity mode closest to the GW frequency and applying the rotating-wave approximation, the Hamiltonian~\cite{Chen:2024aya} is
\begin{subequations}
    \begin{align}
        H &= H_0 + H_{\rm GW},\\
        H_0
        &=
        \omega_c a^\dagger a
        +\omega_q |e\rangle\langle e|\nonumber \\
        & \qquad \qquad +i\lambda_c
        \left(
        a|e\rangle\langle g|
        -a^\dagger|g\rangle\langle e|
        \right),\\
        H_{\rm GW}
        &=
        i\epsilon_c
        \left[
        ae^{i(\omega_g t-\alpha)}
        -a^\dagger e^{-i(\omega_g t-\alpha)}
        \right],
    \end{align}
\end{subequations}
where $\lambda_c$ is the vacuum cavity--qubit coupling and $\epsilon_c$ is the GW-induced cavity-drive amplitude determined by the mode projection in Eq.~\eqref{eqn:detuned_field}. 
The former transfers an excitation coherently between the cavity and qubit, whereas the latter injects the weak external signal into the cavity component.

Because the GW signal is weak, the excitation probability remains much smaller than unity, and it is sufficient to restrict the Hilbert space to the zero- and one-excitation subspaces, $|g,0\rangle,\,|g,1\rangle,\,|e,0\rangle$.
The ground state $|g,0\rangle$ is unchanged, while the one-excitation sector is diagonalized by the cavity--qubit interaction.

After an irrelevant phase redefinition of $|e,0\rangle$ that makes the off-diagonal coupling real, the one-excitation Hamiltonian is a two-level mixing problem.
Writing
\begin{align}
    |\pm\rangle
    =
    C_c^{\pm}|g,1\rangle
    +C_q^{\pm}|e,0\rangle,
\end{align}
the hybrid eigenfrequencies are
\begin{align}
    \omega_\pm
    =
    \frac{\omega_c+\omega_q}{2}
    \pm
    \frac12
    \sqrt{
    (\omega_q-\omega_c)^2
    +
    4|\lambda_c|^2
    },
    \label{eq:hybrid_frequency}
\end{align}
which reduce to the uncoupled cavity and qubit frequencies for $\lambda_c\rightarrow0$. With $\Delta_{qc}\equiv\omega_q-\omega_c$ and $\Omega_R\equiv\sqrt{\Delta_{qc}^2+4|\lambda_c|^2}$, their cavity and qubit fractions can be written as
\begin{align}
    |C_c^{\pm}|^2
    &=
    \frac12\left(1\mp\frac{\Delta_{qc}}{\Omega_R}\right),
    &
    |C_q^{\pm}|^2
    &=
    \frac12\left(1\pm\frac{\Delta_{qc}}{\Omega_R}\right).
    \label{eq:polariton_weights}
\end{align}
At exact resonance the two components are equally mixed and the eigenmodes are separated by the vacuum Rabi splitting $2|\lambda_c|$. 
The GW drive couples $|g,0\rangle$ to a polariton through its cavity fraction $C_c^{\pm}$, while a qubit-population measurement is sensitive to its qubit fraction $C_q^{\pm}$.
Consequently, the resonant cavity enhancement and the qubit readout contrast cannot in general be optimized independently.  
The prescribed-field result is recovered only when the backaction of the qubit on the driven cavity can be neglected.

To include losses, we describe the driven cavity--qubit density matrix by~\cite{Blais:2020wjs}
\begin{align}
    \dot\rho
    ={}&-i[H,\rho]
    +\kappa(\bar n_{\rm th}+1)\mathcal D[a]\rho
    +\kappa\bar n_{\rm th}\mathcal D[a^\dagger]\rho
    \nonumber\\
    &+\gamma_\downarrow\mathcal D[\sigma_-]\rho
    +\gamma_\uparrow\mathcal D[\sigma_+]\rho
    +\frac{\gamma_\phi}{2}\mathcal D[\sigma_z]\rho,
    \label{eq:open_system_master}
\end{align}
where
\begin{align}
    \mathcal D[L]\rho
    \equiv
    L\rho L^\dagger
    -\frac12\{L^\dagger L,\rho\},
    \qquad
    \kappa=\frac{\omega_c}{Q}.
\end{align}
Here $\bar n_{\rm th}=1/(e^{\omega_c/T_{\rm cav}}-1)$ is the cavity thermal occupation, $\gamma_\downarrow$ and $\gamma_\uparrow$ are the qubit relaxation and thermal-excitation rates, and $\gamma_\phi$ is the pure-dephasing rate~\cite{Blais:2020wjs}. 
At low temperature $\gamma_\uparrow$ is suppressed and $\gamma_1\simeq\gamma_\downarrow=1/T_1$, while the qubit coherence-decay rate is
$\gamma_2=\gamma_1/2+\gamma_\phi=1/T_2$.

In the weak-damping dressed-state approximation, and neglecting pure-dephasing-induced transitions between the polariton branches, their population-loss rates are approximately
\begin{align}
    \Gamma_\pm
    \simeq
    \kappa|C_c^\pm|^2
    +\gamma_1|C_q^\pm|^2,
    \label{eq:polariton_loss}
\end{align}
with $\gamma_\phi$ providing additional broadening and loss of phase coherence. 
Neglecting this additional dephasing contribution, the driven response is controlled by susceptibilities of the form $[\omega_g-\omega_\pm+i\Gamma_\pm/2]^{-1}$ rather than by the bare-cavity Lorentzian alone.
In the convention $\kappa=\omega_c/Q$, the stored cavity energy decays on the timescale $1/\kappa=Q/\omega_c$, while the cavity-field amplitude relaxes on the timescale $2/\kappa$. These transient relaxation times are not, by themselves, phase-coherence cutoffs for a continuously driven steady-state signal~\cite{Blais:2020wjs}.

A quantitative detector prediction follows from the measured excited-state population
\begin{align}
    P_e(t)
    =
    \mathrm{Tr}\!\left[
    \left(|e\rangle\langle e|\otimes I_c\right)\rho(t)
    \right],
    \label{eq:open_system_pe}
\end{align}
evaluated for the chosen drive, detuning, interrogation time, and readout protocol.

For a Dicke register the same open-system structure introduces additional constraints. 
Independent qubit relaxation and dephasing contribute dissipators $\sum_i\gamma_1\mathcal D[\sigma_-^{(i)}]$ and $\sum_i(\gamma_\phi/2)\mathcal D[\sigma_z^{(i)}]$. 
If the common lossy cavity is adiabatically eliminated, it generates a collective emission channel proportional to $\mathcal D[J_-]$ and, for a thermally occupied cavity mode, a collective absorption channel proportional to $\mathcal D[J_+]$.
The same collective matrix elements that enhance the coherent signal can then enhance absorption from a thermally occupied cavity mode. 
Moreover, for the half-excited Dicke state used below, a single amplitude-damping event changes the excitation parity and can imitate the parity signature of a signal transition. 
The ideal result $p_{\rm sig}^{({\rm Dicke})}=n_D\delta^2\tau^2$ therefore requires not only $\sqrt{n_D}\delta\tau\ll1$ but also a negligible probability for relaxation, thermal excitation, and dephasing-induced leakage during state preparation, interrogation, and readout. 

The sensitivity curves presented in the main text do not solve this driven open-system problem. 
They correspond to the prescribed-field (no-backaction) benchmark, for which the GW-induced cavity field is assumed to be unaffected by the detector qubits, and they neglect polariton hybridization, cavity loading, and protocol-dependent dissipative errors.
A realistic detector sensitivity must instead be obtained from the driven cavity--qubit master equation above, or from its many-qubit extension including the relevant local and collective dissipative channels, and is left for future work.

\subsection{Collective readout and the physical origin of the enhancement.}
\label{app:Ent_Enh}

\begin{figure}[!ht]
    \centering
     \includegraphics[width=\columnwidth]{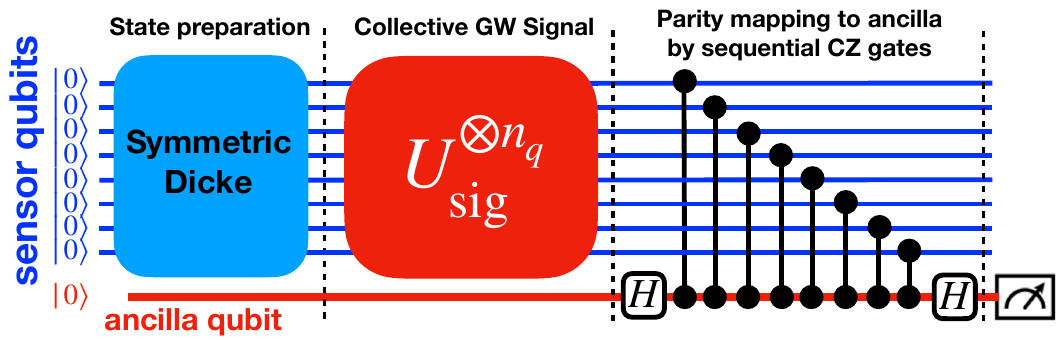}
    \caption{
    Quantum circuit implementing the symmetric Dicke-state protocol. The sensor register is prepared in a symmetric Dicke state, undergoes the signal interaction $U_{\rm sig}^{\otimes n_q}$, and is subsequently mapped onto the ancilla through sequential CZ gates sandwiched between Hadamards for parity readout.
    }
    \label{fig:Dicke_circuit}
\end{figure}

\begin{figure*}[!ht]
    \centering
     \includegraphics[width=0.99\linewidth]{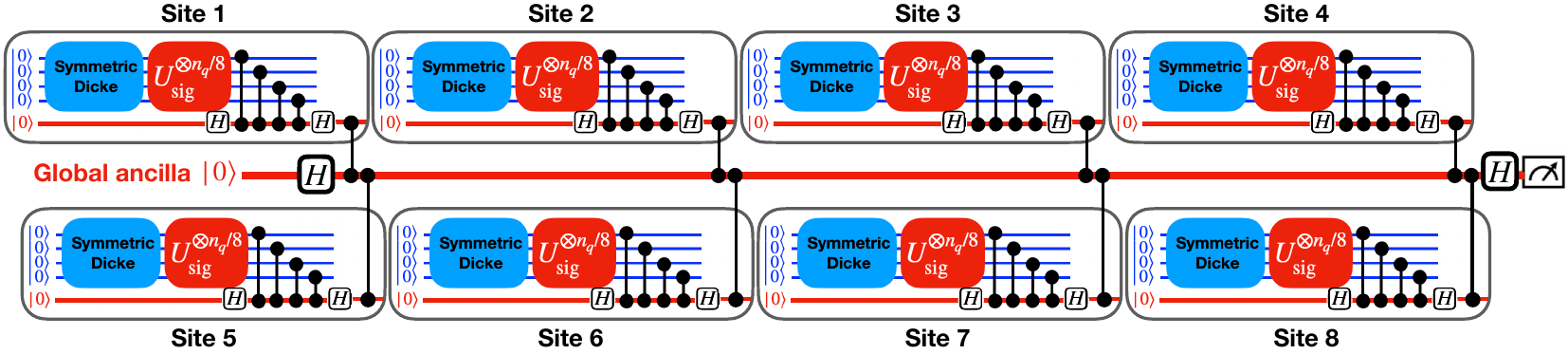}
    \caption{
    Scalable implementation of the local Dicke-state protocol. Each of the eight sensing sites contains $n_q/8$ sensor qubits prepared in a symmetric Dicke state. After the collective GW signal interaction $U_{\rm sig}^{\otimes n_q/8}$, the local parity is mapped onto a local ancilla through sequential CZ gates. The parity information from the eight local ancillae is then transferred to a single global ancilla for the final measurement.
    }
\label{fig:Dicke_circuit_site}
\end{figure*}

The collective protocol enhances this single-qubit response in physically distinct ways, illustrated by the circuits of Figure~\ref{fig:Dicke_circuit}.
For the Dicke-state protocol of Ref.~\cite{Dong:2025mdk}, the sensor register is prepared in a symmetric Dicke state. 
We denote the ground and excited states of each sensor qubit by $|g\rangle$ and $|e\rangle$, respectively. 
The Dicke state with $k$ excited qubits among $n_q$ sensors is
\begin{align}
    |J,m\rangle
    &=
    \binom{n_q}{k}^{-1/2}
    \sum_{\pi}
    \pi\left(
    |e\rangle^{\otimes k}
    |g\rangle^{\otimes(n_q-k)}
    \right),\\
    & \quad \quad \quad \text{where} \quad  J=\frac{n_q}{2}, m = k -\frac{n_q}{2} \nonumber
    \label{eqn:Dicke_state_definition_app}
\end{align}
where the sum runs over all distinct permutations. 

Equivalently,
\begin{align}
    k=J+m,
    \qquad
    n_q-k=J-m .
\end{align}
The collective spin operators are those defined in Eqs.~\eqref{eqn:J_pm_def_counting}--\eqref{eqn:H_int_counting}.  
During one interrogation time $\tau$, the resonant weak GW-induced drive acts on the sensor register as
\begin{align}
    U_{\rm sig}^{\otimes n_q}
    &=
    \exp\left[
    -i\delta\tau
    \left(
    e^{i\alpha}J_+
    +
    e^{-i\alpha}J_-
    \right)
    \right]
    \nonumber\\
    &\simeq
    1
    -i\delta\tau
    \left(
    e^{i\alpha}J_+
    +
    e^{-i\alpha}J_-
    \right),
    \qquad
    \delta\tau\ll1 ,
    \label{eqn:Dicke_Usig_app}
\end{align}
where $\delta$ is the effective single-qubit GW-induced drive amplitude and $\alpha$ is the signal phase.  
Acting on the initial Dicke state, one uses
\begin{align}
        J_+|J,m\rangle
    &=
    \sqrt{(J-m)(J+m+1)}\,|J,m+1\rangle,
    \\
    J_-|J,m\rangle
    &=
    \sqrt{(J+m)(J-m+1)}\,|J,m-1\rangle .        
\end{align}
Therefore
\begin{align}
    |\psi_{\rm sig}\rangle
    &=
    U_{\rm sig}^{\otimes n_q}|J,m\rangle
    \nonumber\\
    &\simeq
    |J,m\rangle
    -i\delta\tau\, e^{i\alpha}
    \sqrt{(J-m)(J+m+1)}
    \,|J,m+1\rangle
    \nonumber\\
    &\hspace{1.2cm}
    -i\delta\tau\, e^{-i\alpha}
    \sqrt{(J+m)(J-m+1)}
    \,|J,m-1\rangle .
    \label{eqn:Dicke_after_signal_app}
\end{align}

We now describe the ancilla-assisted CZ readout used in Figure~\ref{fig:Dicke_circuit}.  The ancilla is initialized in $|g\rangle_A$ and rotated to
\begin{align}
    |+\rangle_A
    =
    \frac{1}{\sqrt2}
    \left(
    |g\rangle_A+|e\rangle_A
    \right).
\end{align}
The CZ gate between the ancilla and the $i$-th sensor is
\begin{align}
    CZ_{Ai}
    =
    |g\rangle_A\langle g|\otimes I_i
    +
    |e\rangle_A\langle e|\otimes Z_i .
\end{align}
For a computational-basis state $|x_r\rangle$ containing $r$ excited sensor qubits,
\begin{align}
    \prod_{i=1}^{n_q}CZ_{Ai}
    \left(
    |+\rangle_A\otimes |x_r\rangle
    \right)
    &=
    \frac{1}{\sqrt2}
    \left(
    |g\rangle_A
    +
    (-1)^r|e\rangle_A
    \right)
    \otimes |x_r\rangle .
\end{align}
Since the Dicke state $|J,m\rangle$ is a symmetric superposition of computational-basis states with the same excitation number $r=J+m$, the same relation holds for Dicke states:
\begin{align}
    \prod_{i=1}^{n_q}CZ_{Ai}
    \left(
    |+\rangle_A\otimes |J,m\rangle
    \right)
    &=
    \frac{1}{\sqrt2}
    \left(
    |g\rangle_A
    +
    (-1)^{r}|e\rangle_A
    \right)
    \otimes |J,m\rangle .
    \label{eqn:CZ_Dicke_parity_app}
\end{align}
After the final Hadamard gate on the ancilla,
\begin{align}
    H_A
    \left[\frac{1}{\sqrt2}
    \left(
    |g\rangle_A+(-1)^r|e\rangle_A
    \right)
    \right]
    =
    \begin{cases}
        |g\rangle_A, & r\ \mathrm{even},\\
        |e\rangle_A, & r\ \mathrm{odd}.
    \end{cases}
    \label{eqn:Dicke_parity_readout_app}
\end{align}
Thus the CZ layer followed by the final Hadamard maps the excitation-number parity of the Dicke register onto the ancilla.
Applying this readout to Eq.~\eqref{eqn:Dicke_after_signal_app}, the initial component $|J,m\rangle$ has excitation number $J+m$, while the signal-induced components $|J,m\pm1\rangle$ have excitation number $J+m\pm1$ and hence the opposite parity.
Hence, up to the trivial choice of whether the signal corresponds to ancilla outcome $|g\rangle_A$ or $|e\rangle_A$, the signal probability is the probability to populate the neighboring Dicke manifolds:
\begin{align}
    p_{\rm sig}^{({\rm Dicke})}
    &=
    \left|
    -i\delta\tau\, e^{i\alpha}
    \sqrt{(J-m)(J+m+1)}
    \right|^2
    \nonumber\\
    &\qquad+
    \left|
    -i\delta\tau\, e^{-i\alpha}
    \sqrt{(J+m)(J-m+1)}
    \right|^2
    \nonumber\\
    &=
    \delta^2\tau^2
    \left[
    2J(J+1)-2m^2
    \right].
\label{eqn:Dicke_probability_app}
\end{align}
valid in the weak-drive regime $\sqrt{2J(J+1)-2m^2} \delta \tau \ll 1$.
Defining the Dicke enhancement factor by
\begin{align}
    n_D
    \equiv
    2J(J+1)-2m^2 ,
    \label{eqn:nD_general_app}
\end{align}
we obtain
\begin{align}
    p_{\rm sig}^{({\rm Dicke})}
    =
    n_D\,\delta^2\tau^2.
\end{align}
where $\sqrt{n_D}\delta \tau \ll 1$.
For the half-excited Dicke state, $m=0$ and $J=n_q/2$, so
\begin{align}
    n_D
    &=
    2J(J+1)
    =
    \frac{n_q^2}{2}+n_q ,                               
\label{eqn:nD_app}
\end{align}
in agreement with Eq.~\eqref{eqn:nD_m0_counting}.  
We note that a single amplitude-damping event during the interrogation also changes the excitation parity and is indistinguishable from a signal transition at the parity level; the plotted Dicke curves are therefore ideal benchmarks that assume no relaxation within a shot.

\begin{table*}[htbp]
    \centering
\renewcommand{\arraystretch}{1.3}
    \resizebox{\textwidth}{!}{%
    \begin{tabular}{lcccccccccccc}
        \hline\hline
        Experiment
        & $f_{\mathrm{TM}_{010}}$ [GHz]
        & $f_{\mathrm{TE}_{212}}$ [GHz]
        & $Q$
        & $t_{\rm int}$
        & $B_0$ [T]
        & $R$ [cm]
        & $L$ [cm]
        & $T_{\rm sys}$ [K]
        & $\eta_{\rm cav,rms}^{\mathrm{TM}_{010}}$
        & ${\cal R}_{q,\rm rms}^{\mathrm{TM}_{010}}$
        & $\eta_{\rm cav,rms}^{\mathrm{TE}_{212}}$
        & ${\cal R}_{q,\rm rms}^{\mathrm{TE}_{212}}$ \\
        \hline
        ADMX~\cite{ADMX:2020hay}
        & 0.57
        & 0.79
        & $3\times10^{4}$
        & $100~\mathrm{s}$
        & 7.6
        & 20
        & 100
        & $0.40$
        & $0.013$
        & $0.025$
        & $0.095$
        & $0.132$ \\
        HAYSTAC~\cite{HAYSTAC:2018rwy}
        & 2.25
        & 3.09
        & $1\times10^{4}$
        & $900~\mathrm{s}$
        & 9.4
        & 5.1
        & 25.4
        & $0.63$
        & $0.013$
        & $0.025$
        & $0.095$
        & $0.132$ \\
        CAPP-12TB~\cite{CAPP:2024dtx}
        & 0.84
        & 1.17
        & $1\times10^{5}$
        & $192~\mathrm{s}$
        & 10.6
        & 13.7
        & 64
        & $0.25$
        & $0.013$
        & $0.026$
        & $0.092$
        & $0.127$ \\
        QUAX~\cite{QUAX:2024fut}
        & 8.49
        & 10.86
        & $1\times10^{5}$
        & $3600~\mathrm{s}$
        & 8.0
        & 1.35
        & 24.6
        & $4.70$
        & $0.005$
        & $0.009$
        & $0.230$
        & $0.318$ \\
        ORGAN~\cite{Quiskamp:2022xur}
        & 7.17
        & 9.85
        & $3.5\times10^{3}$
        & $6000~\mathrm{s}$
        & 11.5
        & 1.6
        & 8.0
        & $6.0$
        & $0.013$
        & $0.025$
        & $0.095$
        & $0.132$ \\
        \hline\hline
    \end{tabular}%
    }
    \caption{
    Representative cavity and operating parameters used for the haloscope benchmarks.
    The eigenfrequencies $f_{\mathrm{TM}_{010}}$ and $f_{\mathrm{TE}_{212}}$ are the corresponding cavity eigenfrequencies computed from the reported geometry via Eq.~\eqref{eqn:res_freq_exact}.
    The last four columns are the orientation averages of Eqs.~\eqref{eqn:eta_sky_rms} and~\eqref{eqn:Rq_sky_rms}, evaluated for a plus-polarized GW with the reported sizes $(R,L)$ of each experiment.
    ADMX and ORGAN share the same aspect ratio $L/R=5$ and therefore give identical averages.
    See the discussion above for the source of $T_{\rm sys}$, which sets both the radiometer and qubit noise (Eqs.~\eqref{eq:cavity_noise_energy} and~\eqref{eqn:p_bkg_cases}).
    }
    \label{tab:haloscope_parameters}
\end{table*}

This is the Dicke superradiant or superabsorbing matrix-element enhancement~\cite{Dicke:1954zz,Higgins:2014cxq,Dong:2025mdk}.  
The two final states $|J,m+1\rangle$ and $|J,m-1\rangle$ are orthogonal and each enters through $|e^{\pm i\alpha}|^2 = 1$, so their probabilities add incoherently and the signal phase $\alpha$ cancels in Eq.~\eqref{eqn:Dicke_probability_app}.
Unlike the GHZ benchmark, no phase-averaging factor appears.
Since $n_D=n_q^2/2+n_q>n_q^2/2$ for $m=0$, the ideal Dicke benchmark is marginally more sensitive than the ideal GHZ protocol at fixed $n_q$, before including preparation, gate, readout, and decoherence errors.

The protocol above assumes that all $n_q$ sensor qubits are prepared in a single symmetric Dicke state. In practice, preparing such a large entangled state is very challenging. A more practical approach is to divide the sensor array into eight independent Dicke registers, one at each ${\rm TE}_{212}$ hot spot, with $n_q/8$ sensor qubits per site. As illustrated in Figure\,\ref{fig:Dicke_circuit_site}, the local parity is first mapped onto a local ancilla and then transferred to a single global ancilla for the final measurement. The corresponding Dicke enhancement becomes
\begin{equation}
n_D^{\rm site} = 8\times\left(\frac{(n_q/8)^2}{2}+n_q/8\right).
\end{equation}
Then the sensitivity would be reduced accordingly. 
For example, for $n_q=800$, the Dicke enhancement is reduced by a factor of
$\frac{n_D}{n_D^{\rm site}}=\frac{8(n_q+2)}{n_q+16}\simeq 7.86,$
while the corresponding strain reach is degraded only by
$\sqrt{\frac{8(n_q+2)}{n_q+16}}\simeq 2.80.$

\section{Haloscope experiment parameters}
\label{app:hal_param}

For each experiment, the temperature entering both noise channels of
Sec.~\ref{sec:Signal_Detection} is the reported system-noise temperature $T_{\rm sys}$ rather than a physical cavity temperature: $T_{\rm sys}$ is substituted directly for the temperature
argument of Eq.~\eqref{eq:cavity_noise_energy} and Eq.~\eqref{eqn:p_bkg_cases}, i.e. the qubit-bath temperature is taken equal to the system noise, $T_q \simeq T_{\rm sys}$.
The five values are: ADMX $0.40~\mathrm{K}$ (JPA-input system noise, range $0.2$–$0.6~\mathrm{K}$)
\cite{ADMX:2020hay}; HAYSTAC $0.63~\mathrm{K}$ ($\approx2.3\times$ the standard quantum limit at
the center frequency) \cite{HAYSTAC:2018rwy}; CAPP-12TB $0.25~\mathrm{K}$ (total system noise,
representative of the $0.15$–$0.4~\mathrm{K}$ run range) \cite{CAPP:2024dtx}; QUAX-LNF
$4.70~\mathrm{K}$ (average HEMT-readout system noise) \cite{QUAX:2024fut}; and ORGAN
$6.0~\mathrm{K}$, for which no single published scalar exists and which we instead compute from
their noise-temperature model at $\beta\approx2$ with an assumed HEMT $T_A\approx12~\mathrm{K}$
\cite{Quiskamp:2022xur}.
At these $T_{\rm sys}$ the qubit thermal background is already close to saturation ($p_{\rm th}\simeq0.44$–$0.49$ for all five benchmarks), so the collective (Dicke) protocols remain the most formal point of comparison among the qubit readouts.

\bibliography{ref}

@article{Kamionkowski:1993fg,
    author = "Kamionkowski, Marc and Kosowsky, Arthur and Turner, Michael S.",
    title = "{Gravitational radiation from first order phase transitions}",
    eprint = "astro-ph/9310044",
    archivePrefix = "arXiv",
    reportNumber = "IASSNS-HEP-93-44, FERMILAB-PUB-93-235-A",
    doi = "10.1103/PhysRevD.49.2837",
    journal = "Phys. Rev. D",
    volume = "49",
    pages = "2837--2851",
    year = "1994"
}

@article{Domcke:2024eti,
    author = "Domcke, Valerie and Ellis, Sebastian A. R. and Kopp, Joachim",
    title = "{Dielectric haloscopes as gravitational wave detectors}",
    eprint = "2409.06462",
    archivePrefix = "arXiv",
    primaryClass = "hep-ph",
    reportNumber = "CERN-TH-2024-151",
    doi = "10.1103/PhysRevD.111.035031",
    journal = "Phys. Rev. D",
    volume = "111",
    number = "3",
    pages = "035031",
    year = "2025"
}

@article{Espinosa:2010hh,
    author = "Espinosa, Jose R. and Konstandin, Thomas and No, Jose M. and Servant, Geraldine",
    title = "{Energy Budget of Cosmological First-order Phase Transitions}",
    eprint = "1004.4187",
    archivePrefix = "arXiv",
    primaryClass = "hep-ph",
    reportNumber = "CERN-PH-TH-2010-027",
    doi = "10.1088/1475-7516/2010/06/028",
    journal = "JCAP",
    volume = "06",
    pages = "028",
    year = "2010"
}

@article{Hindmarsh:2013xza,
    author = "Hindmarsh, Mark and Huber, Stephan J. and Rummukainen, Kari and Weir, David J.",
    title = "{Gravitational waves from the sound of a first order phase transition}",
    eprint = "1304.2433",
    archivePrefix = "arXiv",
    primaryClass = "hep-ph",
    reportNumber = "HIP-2013-07-TH",
    doi = "10.1103/PhysRevLett.112.041301",
    journal = "Phys. Rev. Lett.",
    volume = "112",
    pages = "041301",
    year = "2014"
}

@article{Hindmarsh:2017gnf,
    author = "Hindmarsh, Mark and Huber, Stephan J. and Rummukainen, Kari and Weir, David J.",
    title = "{Shape of the acoustic gravitational wave power spectrum from a first order phase transition}",
    eprint = "1704.05871",
    archivePrefix = "arXiv",
    primaryClass = "astro-ph.CO",
    reportNumber = "HIP-2017-02-TH, HIP-2017-02/TH",
    doi = "10.1103/PhysRevD.96.103520",
    journal = "Phys. Rev. D",
    volume = "96",
    number = "10",
    pages = "103520",
    year = "2017",
    note = "[Erratum: Phys.Rev.D 101, 089902 (2020)]"
}

@article{Roshan:2024qnv,
    author = "Roshan, Rishav and White, Graham",
    title = "{Using gravitational waves to see the first second of the Universe}",
    eprint = "2401.04388",
    archivePrefix = "arXiv",
    primaryClass = "hep-ph",
    doi = "10.1103/RevModPhys.97.015001",
    journal = "Rev. Mod. Phys.",
    volume = "97",
    number = "1",
    pages = "015001",
    year = "2025"
}

@article{Kersten:2024ucm,
    author = {Kersten, J{\"o}rn and Park, Seong Chan and Park, Yeji and Son, Juhoon and Velasco-Sevilla, Liliana},
    title = "{Gravitational waves from a first-order phase transition of the inflaton}",
    eprint = "2412.17278",
    archivePrefix = "arXiv",
    primaryClass = "hep-ph",
    reportNumber = "CQUeST-2024-0752",
    doi = "10.1088/1475-7516/2025/04/053",
    journal = "JCAP",
    volume = "04",
    pages = "053",
    year = "2025"
}

@article{Biswas:2025rzs,
    author = "Biswas, Anirban and Ganguly, Sougata",
    title = "{Probing low scale leptogenesis through gravitational wave}",
    eprint = "2505.01820",
    archivePrefix = "arXiv",
    primaryClass = "hep-ph",
    reportNumber = "CTPU-PTC-25-11",
    doi = "10.1103/zkrc-lrs5",
    journal = "Phys. Rev. D",
    volume = "113",
    number = "3",
    pages = "035028",
    year = "2026"
}

@book{Maggiore:2007ulw,
    author = "Maggiore, Michele",
    title = "{Gravitational Waves. Vol. 1: Theory and Experiments}",
    doi = "10.1093/acprof:oso/9780198570745.001.0001",
    isbn = "978-0-19-171766-6, 978-0-19-852074-0",
    publisher = "Oxford University Press",
    year = "2007"
}

@book{hill2009electromagnetic,
  title={Electromagnetic fields in cavities: deterministic and statistical theories},
  author={Hill, David A},
  year={2009},
  publisher={John Wiley \& Sons}
}

@book{jackson1999classical,
  title={Classical electrodynamics},
  author={Jackson, John David and Fox, Ronald F},
  year={1999},
  publisher={American Association of Physics Teachers}
}

@book{collin1990field,
  title={Field theory of guided waves},
  author={Collin, Robert E},
  year={1990},
  publisher={John Wiley \& Sons}
}

@book{longair2011high,
  title={High energy astrophysics},
  author={Longair, Malcolm S},
  year={2011},
  publisher={Cambridge university press}
}

@article{LIGOScientific:2016aoc,
    author = "Abbott, B. P. and others",
    collaboration = "LIGO Scientific, Virgo",
    title = "{Observation of Gravitational Waves from a Binary Black Hole Merger}",
    eprint = "1602.03837",
    archivePrefix = "arXiv",
    primaryClass = "gr-qc",
    reportNumber = "LIGO-P150914",
    doi = "10.1103/PhysRevLett.116.061102",
    journal = "Phys. Rev. Lett.",
    volume = "116",
    number = "6",
    pages = "061102",
    year = "2016"
}

@article{LISA:2017pwj,
    author = "Amaro-Seoane, Pau and others",
    collaboration = "LISA",
    title = "{Laser Interferometer Space Antenna}",
    eprint = "1702.00786",
    archivePrefix = "arXiv",
    primaryClass = "astro-ph.IM",
    month = "2",
    year = "2017"
}

@article{MAGIS-100:2021etm,
    author = "Abe, Mahiro and others",
    collaboration = "MAGIS-100",
    title = "{Matter-wave Atomic Gradiometer Interferometric Sensor (MAGIS-100)}",
    eprint = "2104.02835",
    archivePrefix = "arXiv",
    primaryClass = "physics.atom-ph",
    reportNumber = "FERMILAB-PUB-21-031-AD-DI-FESS-QIS-T",
    doi = "10.1088/2058-9565/abf719",
    journal = "Quantum Sci. Technol.",
    volume = "6",
    number = "4",
    pages = "044003",
    year = "2021"
}

@article{Bertoldi:2021rqk,
    author = "Bertoldi, Andrea and others",
    title = "{AEDGE: Atomic experiment for dark matter and gravity exploration in space}",
    doi = "10.1007/s10686-021-09701-3",
    journal = "Exper. Astron.",
    volume = "51",
    number = "3",
    pages = "1417--1426",
    year = "2021"
}

@article{Badurina:2019hst,
    author = "Badurina, L. and others",
    title = "{AION: An Atom Interferometer Observatory and Network}",
    eprint = "1911.11755",
    archivePrefix = "arXiv",
    primaryClass = "astro-ph.CO",
    reportNumber = "AION-2019-001, CERN-TH-2019-199",
    doi = "10.1088/1475-7516/2020/05/011",
    journal = "JCAP",
    volume = "05",
    pages = "011",
    year = "2020"
}

@article{NANOGrav:2020bcs,
    author = "Arzoumanian, Zaven and others",
    collaboration = "NANOGrav",
    title = "{The NANOGrav 12.5 yr Data Set: Search for an Isotropic Stochastic Gravitational-wave Background}",
    eprint = "2009.04496",
    archivePrefix = "arXiv",
    primaryClass = "astro-ph.HE",
    doi = "10.3847/2041-8213/abd401",
    journal = "Astrophys. J. Lett.",
    volume = "905",
    number = "2",
    pages = "L34",
    year = "2020"
}

@article{CMB-S4:2020lpa,
    author = "Abazajian, Kevork and others",
    collaboration = "CMB-S4",
    title = "{CMB-S4: Forecasting Constraints on Primordial Gravitational Waves}",
    eprint = "2008.12619",
    archivePrefix = "arXiv",
    primaryClass = "astro-ph.CO",
    reportNumber = "FERMILAB-PUB-20-468-AE-SCD",
    doi = "10.3847/1538-4357/ac1596",
    journal = "Astrophys. J.",
    volume = "926",
    number = "1",
    pages = "54",
    year = "2022"
}

@article{gertsenshtein1962,
  author  = {Gertsenshtein, M. E.},
  title   = {Wave Resonance of Light and Gravitational Waves},
  journal = {Soviet Physics JETP},
  volume  = {14},
  number  = {1},
  pages   = {84--85},
  year    = {1962}
}

@article{Boccaletti:1970pxw,
    author = "Boccaletti, D. and De Sabbata, V. and Fortini, P. and Gualdi, C.",
    title = "{Conversion of photons into gravitons and vice versa in a static electromagnetic field}",
    doi = "10.1007/BF02710177",
    journal = "Nuovo Cim. B",
    volume = "70",
    number = "2",
    pages = "129--146",
    year = "1970"
}

@article{zeldovich1974emgw,
  author  = {Zel'dovich, Ya. B.},
  title   = {Electromagnetic and gravitational waves in a stationary magnetic field},
  journal = {Soviet Physics JETP},
  volume  = {38},
  number  = {4},
  pages   = {652--655},
  year    = {1974}
}

@article{fortini1982fermi,
  title={Fermi normal co-ordinate system and electromagnetic detectors of gravitational waves: I.—calculation of the metric},
  author={Fortini, PL and Gualdi, C},
  journal={Il Nuovo Cimento B (1971-1996)},
  volume={71},
  number={1},
  pages={37--54},
  year={1982},
  publisher={Springer}
}

@article{Chen:2022quj,
    author = "Chen, Shion and Fukuda, Hajime and Inada, Toshiaki and Moroi, Takeo and Nitta, Tatsumi and Sichanugrist, Thanaporn",
    title = "{Detecting Hidden Photon Dark Matter Using the Direct Excitation of Transmon Qubits}",
    eprint = "2212.03884",
    archivePrefix = "arXiv",
    primaryClass = "hep-ph",
    doi = "10.1103/PhysRevLett.131.211001",
    journal = "Phys. Rev. Lett.",
    volume = "131",
    number = "21",
    pages = "211001",
    year = "2023"
}

@article{Chen:2023swh,
    author = "Chen, Shion and Fukuda, Hajime and Inada, Toshiaki and Moroi, Takeo and Nitta, Tatsumi and Sichanugrist, Thanaporn",
    title = "{Quantum Enhancement in Dark Matter Detection with Quantum Computation}",
    eprint = "2311.10413",
    archivePrefix = "arXiv",
    primaryClass = "hep-ph",
    doi = "10.1103/PhysRevLett.133.021801",
    journal = "Phys. Rev. Lett.",
    volume = "133",
    number = "2",
    pages = "021801",
    year = "2024"
}

@article{Chen:2024aya,
    author = "Chen, Shion and Fukuda, Hajime and Inada, Toshiaki and Moroi, Takeo and Nitta, Tatsumi and Sichanugrist, Thanaporn",
    title = "{Search for QCD axion dark matter with transmon qubits and quantum circuit}",
    eprint = "2407.19755",
    archivePrefix = "arXiv",
    primaryClass = "hep-ph",
    doi = "10.1103/PhysRevD.110.115021",
    journal = "Phys. Rev. D",
    volume = "110",
    number = "11",
    pages = "115021",
    year = "2024"
}

@article{Marzlin:1994ia,
    author = "Marzlin, Karl-Peter",
    title = "{Fermi coordinates for weak gravitational fields}",
    eprint = "gr-qc/9403044",
    archivePrefix = "arXiv",
    reportNumber = "KONS-RGKU-94-04",
    doi = "10.1103/PhysRevD.50.888",
    journal = "Phys. Rev. D",
    volume = "50",
    pages = "888--891",
    year = "1994"
}

@article{Rakhmanov:2014noa,
    author = "Rakhmanov, Malik",
    title = "{Fermi-normal, optical, and wave-synchronous coordinates for spacetime with a plane gravitational wave}",
    eprint = "1409.4648",
    archivePrefix = "arXiv",
    primaryClass = "gr-qc",
    doi = "10.1088/0264-9381/31/8/085006",
    journal = "Class. Quant. Grav.",
    volume = "31",
    pages = "085006",
    year = "2014"
}

@article{Berlin:2021txa,
    author = {Berlin, Asher and Blas, Diego and Tito D'Agnolo, Raffaele and Ellis, Sebastian A. R. and Harnik, Roni and Kahn, Yonatan and Sch{\"u}tte-Engel, Jan},
    title = "{Detecting high-frequency gravitational waves with microwave cavities}",
    eprint = "2112.11465",
    archivePrefix = "arXiv",
    primaryClass = "hep-ph",
    reportNumber = "FERMILAB-PUB-21-724-SQMS-T",
    doi = "10.1103/PhysRevD.105.116011",
    journal = "Phys. Rev. D",
    volume = "105",
    number = "11",
    pages = "116011",
    year = "2022"
}

@article{Kim:2025izt,
    author = "Kim, Younggeun and others",
    title = "{Search for high-frequency gravitational waves via reanalysis of cavity axion data}",
    eprint = "2511.17817",
    archivePrefix = "arXiv",
    primaryClass = "hep-ex",
    doi = "10.1103/yqms-lznb",
    journal = "Phys. Rev. D",
    volume = "113",
    number = "7",
    pages = "072015",
    year = "2026"
}

@article{Aggarwal:2025noe,
    author = "Aggarwal, Nancy and others",
    title = "{Challenges and opportunities of gravitational-wave searches above 10 kHz}",
    eprint = "2501.11723",
    archivePrefix = "arXiv",
    primaryClass = "gr-qc",
    reportNumber = "CERN-TH-2025-014, DESY-25-007",
    doi = "10.1007/s41114-025-00060-5",
    journal = "Living Rev. Rel.",
    volume = "28",
    number = "1",
    pages = "10",
    year = "2025"
}

@article{Franciolini:2022htd,
    author = "Franciolini, Gabriele and Maharana, Anshuman and Muia, Francesco",
    title = "{Hunt for light primordial black hole dark matter with ultrahigh-frequency gravitational waves}",
    eprint = "2205.02153",
    archivePrefix = "arXiv",
    primaryClass = "astro-ph.CO",
    doi = "10.1103/PhysRevD.106.103520",
    journal = "Phys. Rev. D",
    volume = "106",
    number = "10",
    pages = "103520",
    year = "2022"
}

@article{Dolgov:2011cq,
    author = "Dolgov, Alexander D. and Ejlli, Damian",
    title = "{Relic gravitational waves from light primordial black holes}",
    eprint = "1105.2303",
    archivePrefix = "arXiv",
    primaryClass = "astro-ph.CO",
    doi = "10.1103/PhysRevD.84.024028",
    journal = "Phys. Rev. D",
    volume = "84",
    pages = "024028",
    year = "2011"
}

@article{Ireland:2023avg,
    author = "Ireland, Aurora and Profumo, Stefano and Scharnhorst, Jordan",
    title = "{Primordial gravitational waves from black hole evaporation in standard and nonstandard cosmologies}",
    eprint = "2302.10188",
    archivePrefix = "arXiv",
    primaryClass = "gr-qc",
    doi = "10.1103/PhysRevD.107.104021",
    journal = "Phys. Rev. D",
    volume = "107",
    number = "10",
    pages = "104021",
    year = "2023"
}

@article{Liebling:2012fv,
    author = "Liebling, Steven L. and Palenzuela, Carlos",
    title = "{Dynamical boson stars}",
    eprint = "1202.5809",
    archivePrefix = "arXiv",
    primaryClass = "gr-qc",
    doi = "10.1007/s41114-023-00043-4",
    journal = "Living Rev. Rel.",
    volume = "26",
    number = "1",
    pages = "1",
    year = "2023"
}

@article{Visinelli:2021uve,
    author = "Visinelli, Luca",
    title = "{Boson stars and oscillatons: A review}",
    eprint = "2109.05481",
    archivePrefix = "arXiv",
    primaryClass = "gr-qc",
    doi = "10.1142/S0218271821300068",
    journal = "Int. J. Mod. Phys. D",
    volume = "30",
    number = "15",
    pages = "2130006",
    year = "2021"
}

@article{Arvanitaki:2014wva,
    author = "Arvanitaki, Asimina and Baryakhtar, Masha and Huang, Xinlu",
    title = "{Discovering the QCD Axion with Black Holes and Gravitational Waves}",
    eprint = "1411.2263",
    archivePrefix = "arXiv",
    primaryClass = "hep-ph",
    doi = "10.1103/PhysRevD.91.084011",
    journal = "Phys. Rev. D",
    volume = "91",
    number = "8",
    pages = "084011",
    year = "2015"
}

@article{Brito:2015oca,
    author = "Brito, Richard and Cardoso, Vitor and Pani, Paolo",
    title = "{Superradiance}: {New Frontiers in Black Hole
Physics}",
    eprint = "1501.06570",
    archivePrefix = "arXiv",
    primaryClass = "gr-qc",
    doi = "10.1007/978-3-319-19000-6",
    journal = "Lect. Notes Phys.",
    volume = "906",
    pages = "pp.1--237",
    year = "2015"
}

@article{Brito:2017zvb,
    author = "Brito, Richard and Ghosh, Shrobana and Barausse, Enrico and Berti, Emanuele and Cardoso, Vitor and Dvorkin, Irina and Klein, Antoine and Pani, Paolo",
    title = "{Gravitational wave searches for ultralight bosons with LIGO and LISA}",
    eprint = "1706.06311",
    archivePrefix = "arXiv",
    primaryClass = "gr-qc",
    doi = "10.1103/PhysRevD.96.064050",
    journal = "Phys. Rev. D",
    volume = "96",
    number = "6",
    pages = "064050",
    year = "2017"
}

@article{Ackley:2020atn,
    author = "Ackley, K. and others",
    title = "{Neutron Star Extreme Matter Observatory: A kilohertz-band gravitational-wave detector in the global network}",
    eprint = "2007.03128",
    archivePrefix = "arXiv",
    primaryClass = "astro-ph.HE",
    doi = "10.1017/pasa.2020.39",
    journal = "Publ. Astron. Soc. Austral.",
    volume = "37",
    pages = "e047",
    year = "2020"
}

@article{Akutsu:2008qv,
    author = "Akutsu, Tomotada and others",
    title = "{Search for a stochastic background of 100-MHz gravitational waves with laser interferometers}",
    eprint = "0803.4094",
    archivePrefix = "arXiv",
    primaryClass = "gr-qc",
    doi = "10.1103/PhysRevLett.101.101101",
    journal = "Phys. Rev. Lett.",
    volume = "101",
    pages = "101101",
    year = "2008"
}

@article{Aggarwal:2020umq,
    author = "Aggarwal, Nancy and Winstone, George P. and Teo, Mae and Baryakhtar, Masha and Larson, Shane L. and Kalogera, Vicky and Geraci, Andrew A.",
    title = "{Searching for New Physics with a Levitated-Sensor-Based Gravitational-Wave Detector}",
    eprint = "2010.13157",
    archivePrefix = "arXiv",
    primaryClass = "gr-qc",
    doi = "10.1103/PhysRevLett.128.111101",
    journal = "Phys. Rev. Lett.",
    volume = "128",
    number = "11",
    pages = "111101",
    year = "2022"
}

@article{Goryachev:2014yra,
    author = "Goryachev, Maxim and Tobar, Michael E.",
    title = "{Gravitational Wave Detection with High Frequency Phonon Trapping Acoustic Cavities}",
    eprint = "1410.2334",
    archivePrefix = "arXiv",
    primaryClass = "gr-qc",
    reportNumber = "Erratum: Phys. Rev. D, 108, 129901(E) (2023)",
    doi = "10.1103/PhysRevD.90.102005",
    journal = "Phys. Rev. D",
    volume = "90",
    number = "10",
    pages = "102005",
    year = "2014",
    note = "[Erratum: Phys.Rev.D 108, 129901 (2023)]"
}

@article{Goryachev:2014nna,
    author = "Goryachev, Maxim and Ivanov, Eugene N. and van Kann, Frank and Galliou, Serge and Tobar, Michael E.",
    title = "{Observation of the Fundamental Nyquist Noise Limit in an Ultra-High $Q$-Factor Cryogenic Bulk Acoustic Wave Cavity}",
    eprint = "1410.4293",
    archivePrefix = "arXiv",
    primaryClass = "physics.ins-det",
    doi = "10.1063/1.4898813",
    journal = "Appl. Phys. Lett.",
    volume = "105",
    pages = "153505",
    year = "2014"
}

@article{Gottardi:2007zn,
    author = "Gottardi, L. and de Waard, A. and Usenko, A. and Frossati, G. and Podt, M. and Flokstra, J. and Bassan, M. and Fafone, V. and Minenkov, Y. and Rocchi, A.",
    title = "{Sensitivity of the spherical gravitational wave detector MiniGRAIL operating at 5 K}",
    eprint = "0705.0122",
    archivePrefix = "arXiv",
    primaryClass = "gr-qc",
    doi = "10.1103/PhysRevD.76.102005",
    journal = "Phys. Rev. D",
    volume = "76",
    pages = "102005",
    year = "2007"
}

@article{Anandan:1982is,
    author = "Anandan, J. and Chiao, R. Y.",
    title = "{GRAVITATIONAL RADIATION ANTENNAS USING THE SAGNAC EFFECT}",
    doi = "10.1007/BF00756213",
    journal = "Gen. Rel. Grav.",
    volume = "14",
    pages = "515--521",
    year = "1982"
}

@article{Asztalos2010,
  title = {SQUID-Based Microwave Cavity Search for Dark-Matter Axions},
  author = {Asztalos, S. J. and others},
  journal = {Phys. Rev. Lett.},
  volume = {104},
  pages = {041301},
  year = {2010},
  doi = {10.1103/PhysRevLett.104.041301}
}

@article{ADMX:2018gho,
    author = "Du, N. and others",
    collaboration = "ADMX",
    title = "{A Search for Invisible Axion Dark Matter with the Axion Dark Matter Experiment}",
    eprint = "1804.05750",
    archivePrefix = "arXiv",
    primaryClass = "hep-ex",
    reportNumber = "FERMILAB-PUB-18-101-AD-AE",
    doi = "10.1103/PhysRevLett.120.151301",
    journal = "Phys. Rev. Lett.",
    volume = "120",
    number = "15",
    pages = "151301",
    year = "2018"
}

@article{ADMX:2020hay,
    author = "Bartram, C. and others",
    collaboration = "ADMX",
    title = "{Axion dark matter experiment: Run 1B analysis details}",
    eprint = "2010.06183",
    archivePrefix = "arXiv",
    primaryClass = "astro-ph.CO",
    reportNumber = "FERMILAB-PUB-20-573-AD-AE",
    doi = "10.1103/PhysRevD.103.032002",
    journal = "Phys. Rev. D",
    volume = "103",
    number = "3",
    pages = "032002",
    year = "2021"
}

@article{ADMX:2021nhd,
    author = "Bartram, C. and others",
    collaboration = "ADMX",
    title = "{Search for Invisible Axion Dark Matter in the 3.3{\textendash}4.2{\,}{\,}{\ensuremath{\mu}}eV Mass Range}",
    eprint = "2110.06096",
    archivePrefix = "arXiv",
    primaryClass = "hep-ex",
    reportNumber = "FERMILAB-PUB-21-774-DI-PPD-SQMS",
    doi = "10.1103/PhysRevLett.127.261803",
    journal = "Phys. Rev. Lett.",
    volume = "127",
    number = "26",
    pages = "261803",
    year = "2021"
}

@article{HAYSTAC:2018rwy,
    author = "Zhong, L. and others",
    collaboration = "HAYSTAC",
    title = "{Results from phase 1 of the HAYSTAC microwave cavity axion experiment}",
    eprint = "1803.03690",
    archivePrefix = "arXiv",
    primaryClass = "hep-ex",
    doi = "10.1103/PhysRevD.97.092001",
    journal = "Phys. Rev. D",
    volume = "97",
    number = "9",
    pages = "092001",
    year = "2018"
}

@article{CAPP:2024dtx,
    author = "Ahn, Saebyeok and others",
    collaboration = "CAPP",
    title = "{Extensive Search for Axion Dark Matter over 1~GHz with CAPP{\textquoteright}S Main Axion Experiment}",
    eprint = "2402.12892",
    archivePrefix = "arXiv",
    primaryClass = "hep-ex",
    doi = "10.1103/PhysRevX.14.031023",
    journal = "Phys. Rev. X",
    volume = "14",
    number = "3",
    pages = "031023",
    year = "2024"
}

@article{CAPP:2020utb,
    author = "Kwon, Ohjoon and others",
    collaboration = "CAPP",
    title = "{First Results from an Axion Haloscope at CAPP around 10.7  $\mu$eV}",
    eprint = "2012.10764",
    archivePrefix = "arXiv",
    primaryClass = "hep-ex",
    doi = "10.1103/PhysRevLett.126.191802",
    journal = "Phys. Rev. Lett.",
    volume = "126",
    number = "19",
    pages = "191802",
    year = "2021"
}

@article{McAllister:2017lkb,
    author = "McAllister, Ben T. and Flower, Graeme and Kruger, Justin and Ivanov, Eugene N. and Goryachev, Maxim and Bourhill, Jeremy and Tobar, Michael E.",
    collaboration = "ORGAN",
    title = "{The ORGAN Experiment: An axion haloscope above 15 GHz}",
    eprint = "1706.00209",
    archivePrefix = "arXiv",
    primaryClass = "physics.ins-det",
    doi = "10.1016/j.dark.2017.09.010",
    journal = "Phys. Dark Univ.",
    volume = "18",
    pages = "67--72",
    year = "2017"
}

@article{Quiskamp:2022xur,
    author = "Quiskamp, Aaron P. and McAllister, Ben T. and Altin, Paul
              and Ivanov, Eugene N. and Goryachev, Maxim and Tobar, Michael E.",
    collaboration = "ORGAN",
    title = "{Direct search for dark matter axions excluding ALP cogenesis in the
              63- to 67-$\mu$eV range with the ORGAN experiment}",
    journal = "Sci. Adv.",
    volume = "8",
    number = "27",
    pages = "eabq3765",
    year = "2022",
    eprint = "2203.12152",
    archivePrefix = "arXiv",
    primaryClass = "hep-ex",
    doi = "10.1126/sciadv.abq3765"
}

@article{Alesini:2019ajt,
    author = "Alesini, D. and others",
    title = "{Galactic axions search with a superconducting resonant cavity}",
    eprint = "1903.06547",
    archivePrefix = "arXiv",
    primaryClass = "physics.ins-det",
    doi = "10.1103/PhysRevD.99.101101",
    journal = "Phys. Rev. D",
    volume = "99",
    number = "10",
    pages = "101101",
    year = "2019"
}

@article{QUAX:2024fut,
    author = "Rettaroli, A. and others",
    collaboration = "QUAX",
    title = "{Search for axion dark matter with the QUAX{\textendash}LNF tunable haloscope}",
    eprint = "2402.19063",
    archivePrefix = "arXiv",
    primaryClass = "physics.ins-det",
    reportNumber = "FERMILAB-PUB-24-0511-SQMS-V",
    doi = "10.1103/PhysRevD.110.022008",
    journal = "Phys. Rev. D",
    volume = "110",
    number = "2",
    pages = "022008",
    year = "2024"
}

@article{Jeong:2021dij,
    author = "Jeong, Junu and Youn, SungWoo and Bae, Sungjae and Kim, Dongok and Kim, Younggeun and Semertzidis, Yannis K.",
    title = "{Analytical considerations for optimal axion haloscope design}",
    eprint = "2108.00608",
    archivePrefix = "arXiv",
    primaryClass = "hep-ex",
    doi = "10.1088/1361-6471/ac58b4",
    journal = "J. Phys. G",
    volume = "49",
    number = "5",
    pages = "055201",
    year = "2022"
}

@article{Dong:2025mdk,
    author = "Dong, Zhongtian and Kim, Doojin and Kong, Kyoungchul and Park, Myeonghun and Alcaraz, Miguel A. Soto",
    title = "{Quantum Sensing Radiative Decays of Neutrinos and Dark Matter Particles}",
    eprint = "2508.09139",
    archivePrefix = "arXiv",
    primaryClass = "hep-ph",
    month = "8",
    year = "2025"
}

@article{Dicke:1954zz,
    author = "Dicke, R. H.",
    title = "{Coherence in Spontaneous Radiation Processes}",
    doi = "10.1103/PhysRev.93.99",
    journal = "Phys. Rev.",
    volume = "93",
    pages = "99--110",
    year = "1954"
}

@article{Higgins:2014cxq,
    author = "Higgins, K. D. B. and Benjamin, S. C. and Stace, T. M. and Milburn, G. J. and Lovett, B. W. and Gauger, E. M.",
    title = "{Superabsorption of light via quantum engineering}",
    eprint = "1306.1483",
    archivePrefix = "arXiv",
    primaryClass = "quant-ph",
    doi = "10.1038/ncomms5705",
    month = "2",
    year = "2014"
}

@article{He:2025ovo,
    author = "He, Ping and Shu, Jing and Xu, Bin and Xu, Jincheng",
    title = "{Symmetric Dicke States as Optimal Probes for Wave-Like Dark Matter}",
    eprint = "2512.14821",
    archivePrefix = "arXiv",
    primaryClass = "hep-ph",
    month = "12",
    year = "2025"
}

@article{Kim:2023bwr,
    author = "Kim, Youngseok and others",
    title = "{Evidence for the utility of quantum computing before fault tolerance}",
    doi = "10.1038/s41586-023-06096-3",
    journal = "Nature",
    volume = "618",
    number = "7965",
    pages = "500--505",
    year = "2023"
}

@online{IBMQuantumHardware,
  author  = {{IBM}},
  title   = {{IBM Quantum Hardware}},
  year    = {2026},
  url     = {https://www.ibm.com/quantum/hardware},
  urldate = {\today}
}

@article{Blais:2020wjs,
    author = "Blais, Alexandre and Grimsmo, Arne L. and Girvin, S. M. and Wallraff, Andreas",
    title = "{Circuit quantum electrodynamics}",
    eprint = "2005.12667",
    archivePrefix = "arXiv",
    primaryClass = "quant-ph",
    doi = "10.1103/RevModPhys.93.025005",
    journal = "Rev. Mod. Phys.",
    volume = "93",
    number = "2",
    pages = "025005",
    year = "2021"
}

@article{Krause:2021llk,
    author = "Krause, J. and Dickel, C. and Vaal, E. and Vielmetter, M. and Feng, J. and Bounds, R. and Catelani, G. and Fink, J. M. and Ando, Yoichi",
    title = "{Magnetic Field Resilience of Three-Dimensional Transmons with Thin-Film Al/AlOx/Al Josephson Junctions Approaching 1 T}",
    eprint = "2111.01115",
    archivePrefix = "arXiv",
    primaryClass = "quant-ph",
    doi = "10.1103/PhysRevApplied.17.034032",
    journal = "Phys. Rev. Applied",
    volume = "17",
    number = "3",
    pages = "034032",
    year = "2022"
}

@article{Gunzler:2025spt,
    author = {G{\"u}nzler, S. and others},
    title = "{Spin environment of a superconducting qubit in high magnetic fields}",
    eprint = "2501.03661",
    archivePrefix = "arXiv",
    primaryClass = "quant-ph",
    doi = "10.1038/s41467-025-65528-y",
    journal = "Nature Commun.",
    volume = "16",
    number = "1",
    pages = "9564",
    year = "2025"
}

@article{PhysRevApplied.15.054001,
  title = {Magnetic-Field-Compatible Superconducting Transmon Qubit},
  author = {Kringh\o{}j, A. and Larsen, T. W. and Erlandsson, O. and Uilhoorn, W. and Kroll, J.G. and Hesselberg, M. and McNeil, R.P.G. and Krogstrup, P. and Casparis, L. and Marcus, C.M. and Petersson, K.D.},
  journal = {Phys. Rev. Appl.},
  volume = {15},
  issue = {5},
  pages = {054001},
  numpages = {8},
  year = {2021},
  month = {May},
  publisher = {American Physical Society},
  doi = {10.1103/PhysRevApplied.15.054001},
  url = {https://link.aps.org/doi/10.1103/PhysRevApplied.15.054001}
}

@article{Kono:2017ynw,
    author = "Kono, S. and Koshino, K. and Tabuchi, Y. and Noguchi, A. and Nakamura, Y.",
    title = "{Quantum non-demolition detection of an itinerant microwave photon}",
    eprint = "1711.05479",
    archivePrefix = "arXiv",
    primaryClass = "quant-ph",
    doi = "10.1038/s41567-018-0066-3",
    journal = "Nature Phys.",
    volume = "14",
    number = "6",
    pages = "546--549",
    year = "2018"
}

\end{document}